\documentclass[runningheads, envcountsame, a4paper]{llncs}
\usepackage{epstopdf}
\usepackage[hyphens]{url}
\usepackage{microtype}

\usepackage{amssymb,amsmath}
\usepackage[toc,page]{appendix}
\usepackage{placeins}
\usepackage{graphicx}
\usepackage{algorithm,algorithmic}
\usepackage{tikz}
\usetikzlibrary{plotmarks}
\usetikzlibrary{automata} 
\usetikzlibrary[automata] 
\usetikzlibrary{decorations}
\usepackage{xifthen}
\usepackage{pstricks,pst-node,pst-tree}
\usepackage{fancybox}
\usetikzlibrary{shapes.arrows,chains}
\usepackage[all]{xy}
\usepackage{bibentry}
\usepackage{graphics}
\usepackage{xspace,aeguill,ae}
\usepackage{verbatim,mathrsfs,enumerate,url}
\usepackage{nag}
\usepackage{wrapfig}

\begin{document}
\newcommand{\setPayoff}{P}
\newcommand{\fctPayoff}{f}
\newcommand{\maxCol}{C}
\newcommand{\fctCol}{c}
\newcommand{\Infrho}{in\!f}
\newcommand{\Sub}[2]{{#1}_{|{#2}}} 
\newcommand{\Hist}{Hist}
\newcommand{\Plays}{Plays}
\newcommand{\out}[1]{\langle #1 \rangle}
\newcommand{\LimSup}{{\sf LimSup}}
\newcommand{\LimInf}{{\sf LimInf}}
\newcommand{\Sup}{{\sf Sup}}
\newcommand{\Inf}{{\sf Inf}}
\newcommand{\Disc}{{\sf Disc}}
\newcommand{\WMP}{\sf {WMP}}
\newcommand{\InfMP}{{\sf \underline{MP}}}
\newcommand{\SupMP}{{\sf \overline{MP}}}
\newcommand{\limAvgInf}{\liminf\limits_{n \to \infty} \frac{1}{n} \sum\limits_{k=0}^{n-1}}
\newcommand{\limAvgSup}{\limsup\limits_{n \to \infty} \frac{1}{n} \sum\limits_{k=0}^{n-1}}
\newcommand{\val}{{val}}

\title{Computer aided synthesis:\\ a game-theoretic approach}
\author{V\'eronique Bruy\`ere}
\institute{Computer Science Department, University of Mons \\20 Place du Parc, B-7000-Mons, Belgium \\\email{Veronique.Bruyere@umons.ac.be}}

\titlerunning{Computer aided synthesis: a game-theoretic approach}
\authorrunning{V. Bruy\`ere}
\toctitle{Computer aided synthesis: a game-theoretic approach}
\tocauthor{V\'eronique~Bruy\`ere}

\date{}

\maketitle
\setcounter{footnote}{0}

\begin{abstract}
In this invited contribution, we propose a comprehensive introduction to game theory applied in computer aided synthesis. In this context, we give some classical results on two-player zero-sum games and then on multi-player non zero-sum games. The simple case of one-player games is strongly related to automata theory on infinite words. All along the article, we focus on general approaches to solve the studied problems, and we provide several illustrative examples as well as intuitions on the proofs.
\keywords{Games played on graphs, Boolean objective, quantitative objective, winning strategy, Nash equilibrium, synthesis.}
\end{abstract}

\section{Introduction}

\emph{Game theory} is a well-developed branch of mathematics that is applied to various domains like economics, biology, computer science, etc. It is the study of mathematical models of interaction and conflict between individuals and the understanding of their decisions assuming that they are rational~\cite{Osborne94,vonNeumannMorgenstern44}. 

The last decades have seen a lot of research on algorithmic questions in game theory motivated by problems from \emph{computer aided synthesis}. 
One important line of research is concerned with \emph{reactive systems} that must continuously react to the uncontrollable events produced by the environment in which they evolve. A \emph{controller} of a reactive system indicates which actions it has to perform to satisfy a certain objective against any behavior of the environment. An example in air traffic management is the autopilot that controls the speed of the plane, but have no control on the weather conditions. Such a situation can be modeled by a \emph{two-player game played on a graph}: the system and the environment are the two players, the vertices of the graph model the possible configurations, the infinite paths in the graph model all the continuous interactions between the system and the environment. In this game, the system wants to achieve a certain objective while the environment tries to prevent it to do so. The objectives of the two players are thus \emph{antagonistic} and we speak of \emph{zero-sum} games. In this framework, checking whether the system is able to achieve its objective reduces to the existence of a \emph{winning strategy} in the corresponding game, and building a controller reduces to computing such a strategy~\cite{2001automata}. Whether such a controller can be automatically designed from the objective is known as the \emph{synthesis problem}.

Another, more recent, line of research is concerned with the modelization and the study of \emph{complex systems}. Instead of the simple situation of a system embedded in a hostile environment, we are faced with systems/environments formed of several components each of them with their own objectives that are not necessarily conflicting. Imagine the situation of several users behind their computers on a shared network. In this case, we use the model of \emph{multi-player non zero-sum games played on graphs}: the components are the different players, each of them aiming at satisfying his objective. In this context, the synthesis problem is a little different: winning strategies are no longer appropriate and are replaced by the concept of \emph{equilibrium}, that is, a strategy profile where no player has an incentive to deviate~\cite{GU08}. Different kinds of equilibria have been investigated among which the famous notion of \emph{Nash equilibrium}~\cite{Nas50}.

A lot of study has been done about \emph{Boolean} objectives, in particular about the class of \emph{$\omega$-regular} objectives, like avoiding a deadlock, always granting a request, etc~\cite{2001automata}. An infinite path in the game graph is either winning or losing depending on whether the objective is satisfied or not. To allow richer objectives, such as minimizing the energy consumption or guaranteeing a limited response time to a request, existing models have been enriched with quantitative aspects in a way to associate a payoff (or a cost) to all paths in the game graph~\cite{LaurentDoyen}. In this setting, we speak of \emph{quantitative} objectives, and a classical \emph{decision problem} in two-player zero-sum games is whether there exists a winning strategy for the system that ensures a payoff satisfying some given \emph{constraints} no matter how the environment behaves. For instance we would like an energy consumption lying within a certain given interval. The same kind of question is also considered for multi-player non zero-sum games, that is, whether there exists an equilibrium such that the payoff of each player satisfies the constraints. 

Decidability of those problems is not enough. Indeed in case of positive answer, it is important to know the exact \emph{complexity class} of the problem and how complex are the strategies used to solve it. Given past interactions between the players, a strategy for a player indicates the next action he has to perform. The \emph{amount of memory} on those past interactions is one of the ways to express the complexity of the strategy. The simplest strategies are those that require no memory at all. When all these characteristics are known and indicate practical applicability of the models, the final step is the \emph{implementation} of the solving strategies into a program (like for instance a controller for a reactive system) by using adequate data structures and possibly heuristics.

In this article, we propose a \emph{comprehensive introduction} to classical algorithmic solutions to the synthesis problem for two-player zero-sum games and for multi-player non zero-sum games. A complementary survey can be found in~\cite{BrenguierCHPRRS16}, and detailed expositions in the case of Boolean objectives are provided in~\cite{2001automata,GU08}. We study the existence of winning strategies (in two-player zero-sum games) and equilibria (in multi-player non zero-sum games) satisfying some given constraints, in particular the complexity class of the decision problem and the memory required for the related strategies. We provide several illustrative examples as well as intuitions on some proofs. We do not intend to present an exhaustive survey, but rather focus on some lines of research, with an emphasis on \emph{general approaches}. In particular, we only consider \emph{(i)} \emph{turned-based} (and not concurrent) games such that the players choose their actions in a turned-based way (and not concurrently), \emph{(ii)} \emph{deterministic} (and not stochastic) games such that their edges are deterministic and not labeled by probabilities \emph{(iii)} \emph{pure} (and not randomized) strategies such that the next action is chosen in a deterministic way (and not according to a probability distribution). 

Our approach is as follows. We begin with a general definition of game that includes the class of games with Boolean objectives and the class of games with quantitative objectives. For two-player zero-sum games, we present a criterium~\cite{GimbertZ04} that implies, for several large families of games, the existence of memoryless winning strategies ensuring a payoff satisfying some given constraints. For non zero-sum multi-player games, we present a characterization of plays  (used for instance in~\cite{BMR14,UmmelsW11}) that are the outcome of a Nash equilibrium. The existence of Nash equilibrium in many different families of games is derived from this characterization, as well as results on the existence of a Nash equilibrium satisfying some constraints. We also present two other well-studied equilibria: the secure equilibria~\cite{CHJ06} and the subgame perfect equilibria~\cite{selten}. For the studied decision problems, in addition to the results derived from our general approaches, we provide in this survey an overview of known results for games with Boolean and quantitative objectives.

The article is organized in the following way. In Section~\ref{sec:prelim}, we introduce the concepts of game and strategy, we then present the studied decision problems, and we finally recall the Boolean and quantitative objectives that are classically studied. In Section~\ref{sec:two} devoted to two-player zero-sum games, we begin with the simple case of one-player games, and show how the decision problems are connected to problems in automata theory and numeration systems. We then present the general criterium mentioned before, and then the solutions to the decision problems for the classes of games with Boolean and quantitative objectives. Finally, we present several recent extensions of those classes of games, where for instance the single objective is replaced by a Boolean intersection of several objectives. The case of multi-player non zero-sum games is investigated in Section~\ref{sec:multi} by starting with the characterization of outcomes of Nash equilibrium. Derived results on the existence of Nash equilibrium (under some given constraints) are then detailed, followed by a study of other kinds of equilibria like secure and subgame perfect equilibria. We provide a short conclusion in Section~\ref{sec:conclusion}.

\section{Terminology and studied problems} \label{sec:prelim}

We consider multi-player turn-based games played on finite directed graphs. The set of vertices are partitioned among the different players. A play is an infinite sequence of vertices obtained by moving an imaginary pebble from vertex to vertex according to existing edges. The owner of the current vertex decides what is the next move of the pebble according to some strategy. Each player follows a strategy in a way to achieve a certain objective. This objective depends on a preference relation that the player has on the payoffs assigned to plays. In this section, we introduce all these notions and state the problems studied in this article.

\subsection{Preliminaries}

\subsubsection{Games} We begin with the notions of arena and game. 

\begin{definition} \label{def:game}
An \emph{arena} is a tuple $A = (\Pi, V, (V_i)_{i \in \Pi}, E)$ where: 

\begin{itemize}
  \item $\Pi $ is a finite set of \emph{players}, 
  \item $V$ is a finite set of \emph{vertices} and $E \subseteq V \times V$ is a set of \emph{edges}, such that each vertex has at least one outgoing edge\footnote{This condition guarantees that there is no deadlock. It can be assumed w.l.o.g. for all the problems considered in this article.},
  \item $(V_i)_{i \in \Pi}$ is a partition of $V$, where $V_i$ is the set of vertices owned\footnote{We also say that player~$i$ \emph{controls} the vertices of $V_i$.} by player $i \in \Pi$. 
\end{itemize}
\end{definition}

A \emph{play} is an infinite sequence $\rho = \rho_0 \rho_1 \ldots \in V^\omega$ of vertices such that $(\rho_k, \rho_{k + 1}) \in E$ for all $k \in \mathbb N$. \emph{Histories} are finite sequences $h = h_0 \ldots h_n \in V^*$ defined in the same way. We often use notation $hv$ to mention the last vertex $v \in V$ of the history. 
The set of plays is denoted by $\Plays$ and the set of non empty histories (resp. ending with a vertex in $V_i$) by $\Hist$ (resp. by $\Hist_i$).
A \emph{prefix} (resp. \emph{suffix}) of a play $\rho = \rho_0\rho_1 \ldots$ is a finite sequence $\rho_{\leq n} = \rho_0 \dots \rho_n$  (resp. infinite sequence $\rho_{\geq n} = \rho_n \rho_{n+1} \ldots$). 
We often use notation $h\rho$ for a play of which history $h$ is prefix. 
Given a play $\rho$, we denote by $\Infrho(\rho)$ the set of vertices visited infinitely often by $\rho$. We say that $\rho$ is a \emph{lasso} if it is equal to  $hg^\omega$ with $h,g$ being two histories. This lasso is called \emph{simple} if $hg$ has no repeated vertices.

\begin{definition}  \label{def:payoff}
A \emph{game} $G$ is an arena $A = (\Pi, V, (V_i)_{i \in \Pi}, E)$ such that each player~$i$ has:
\begin{itemize}
\item a \emph{payoff function} $f_i~: \Plays \to \setPayoff_i$ where $P_i$ is a set of \emph{payoffs},
\item a \emph{preference relation} $\prec_i$ $\subseteq \setPayoff_i \times \setPayoff_i$ on his set of payoffs.
\end{itemize}
\end{definition}

A \emph{preference} relation $\prec_i$ is a strict total order\footnote{that is, an irreflexive, transitive and total binary relation.}. It allows player~$i$ to compare two plays $\rho, \rho' \in \Plays$ with respect to their payoffs: $f_i(\rho) \prec_i f_i(\rho')$ means that player~$i$ prefers $\rho'$ to $\rho$.  
Given $p, p' \in \setPayoff_i$, we write $p \preceq_i p'$ when $p \prec_i p'$ or $p = p'$; notice that $p \nprec_i p'$ iff $p' \preceq_i p$ since $\prec_i$ is total. 

A payoff function $\fctPayoff_i$ is \emph{prefix-independent} if $\fctPayoff_i(h\rho) = \fctPayoff_i(\rho)$ for all $h\rho \in \Plays$. It is \emph{prefix-linear} if for all $h\rho, h\rho' \in \Plays$,
\begin{eqnarray}
\fctPayoff_i(\rho) \preceq_i \fctPayoff_i(\rho') &\Rightarrow& \fctPayoff_i(h \rho) \preceq_i \fctPayoff_i(h \rho'), and \label{eq:1} \\
\fctPayoff_i(\rho) \prec_i \fctPayoff_i(\rho') &\Rightarrow& \fctPayoff_i(h \rho) \prec_i \fctPayoff_i(h \rho').\label{eq:2}
\end{eqnarray}
Any prefix-independent function $f_i$ is prefix-linear.

When an initial vertex $v_0 \in V$ is fixed, we call $(G, v_0)$ an \emph{initialized} game. In this case, plays and histories are supposed to start in $v_0$, and we then use notations $\Plays(v_0)$, $\Hist(v_0)$, and $\Hist_i(v_0)$ (instead of $\Plays$, $\Hist$, and $\Hist_i$). 

\begin{figure}[ht!]
\begin{center}
\begin{tikzpicture}[initial text=,auto, node distance=2cm, shorten >=1pt] 

\node[state, scale=0.6]              (1)                     {$v_0$};
\node[state, rectangle, scale=0.6]   (2)    [right=of 1]     {$v_1$};
\node[state, scale=0.6]              (3)    [left=of 1]      {$v_2$};
\node[state, scale=0.6]              (4)    [right=of 2]     {$v_3$};

\path[->] (1) edge [bend right=25]          node[below, scale=0.7, black]    {}   (2)
             edge                                node[above, scale=0.7]           {}   (3)

         (2) edge  [bend left=-25]               node[above, scale=0.7]           {}   (1)
             edge  []                        node[above, scale=0.7, black]    {}   (4)

         (3) edge  [loop above]            node[midway, scale=0.7, black]   {}   ()

         (4) edge  [loop above]            node[midway, scale=0.7, black]   {}   ();

\end{tikzpicture}
\end{center}
\caption{A two-player game with payoff functions $f = f_1 = f_2$, preference relations $p_1 \prec_1 p_2 \prec_1 p_3$ and $p_2 \prec_2 p_3 \prec_2 p_1$, such that $\fctPayoff((v_0v_1)^\omega) = p_1$, $\fctPayoff(v_2^\omega) = p_2$, and $\fctPayoff(v_3^\omega) = p_3$}
\label{fig:gameAbstract}
\end{figure}
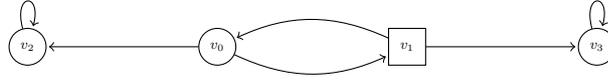

\begin{example} \label{ex:first-ex} 
Consider the initialized two-player game $(G, v_0)$ in Figure~\ref{fig:gameAbstract} such that player~$1$ (resp. player~$2$) controls vertices $v_0, v_2, v_3$ (resp. vertex $v_1$).\footnote{In all examples of this article, circle (resp. square) vertices are controlled by player~$1$ (resp. player~$2$).} Both players use the same set $\setPayoff$ of payoffs equal to $\{p_1, p_2, p_3 \}$, and the same payoff function $\fctPayoff$ that is prefix-independent: $\fctPayoff((v_0v_1)^\omega) = p_1$, $\fctPayoff(v_2^\omega) = p_2$, and $\fctPayoff(v_3^\omega) = p_3$. The preference relation for player~$1$ (resp. player~$2$) is $p_1 \prec_1 p_2 \prec_1 p_3$ (resp. $p_2 \prec_2 p_3 \prec_2 p_1$).
\end{example}

\subsubsection{Strategies}

Let $(G, v_0)$ be an initialized game. A \emph{strategy} $\sigma_i$ for player~$i$ in $(G,v_0)$ is a function $\sigma_i: \Hist_i(v_0) \to V$ assigning to each history $hv \in \Hist_i(v_0)$ a vertex $v' = \sigma_i(hv)$ such that $(v, v') \in E$. Thus $\sigma_i(hv)$ is the next vertex chosen by player~$i$ (that controls vertex $v$) after history $hv$ has been played. A play $\rho \in \Plays(v_0)$ is \emph{consistent} with $\sigma_i$ if $\rho_{n+1} = \sigma_i(\rho_{\leq n})$ for all $n$ such that $\rho_n \in V_i$.

A strategy $\sigma_i$ for player $i$ is \emph{positional} if it only depends on the last vertex of the history, i.e., $\sigma_i(hv) = \sigma_i(v)$ for all $hv \in \Hist_i(v_0)$. More generally, it is \emph{finite-memory} if $\sigma_i(hv)$ needs only a finite information out of the history $hv$. This is possible with a finite-state machine that keeps track of histories of plays. The strategy chooses the next vertex depending on the current state of the machine and the current vertex in the game.\footnote{This informal definition is enough for this survey. See for instance~\cite{2001automata} for a definition.}
The previous definition of positional strategy $\sigma_i$ for player~$i$ is given for an initialized game $(G, v_0)$. We call it \emph{uniform} if it is defined for all $hv \in \Hist_i$ (instead of $\Hist_i(v_0)$), that is, when $\sigma_i$ is a positional strategy in all initialized games $(G,v)$, $v \in V$.
 
A \emph{strategy profile} is a tuple $(\sigma_i)_{i \in \Pi}$ of strategies, where each $\sigma_i$ is a strategy of player~$i$. It is called \emph{positional} (resp. \emph{uniform}, \emph{finite-memory}) if all $\sigma_i$, $i \in \Pi$, are positional (resp. uniform, finite-memory). 
Given an initial vertex $v_0$, such a strategy profile determines a unique play of $(G, v_0)$ that is consistent with all strategies $\sigma_i$. This play is called the \emph{outcome} of $(\sigma_i)_{i \in \Pi}$ in $(G,v_0)$ and is denoted by $\out{(\sigma_i)_{i \in \Pi}}_{v_0}$.

\addtocounter{example}{-1}
\begin{example}[continued]
An example of strategy profile $(\sigma_1,\sigma_2)$ in $(G,v_0)$ is the following one:
\begin{itemize}
\item the positional strategy $\sigma_2$ for player~$2$ is defined such that $\sigma_2(hv_1) = v_3$ for all $hv_1 \in \Hist(v_0)$,
\item the finite-memory strategy $\sigma_1$ for player~$1$ is defined such that $\sigma_1(v_0) = v_1$ and $\sigma_1(hv_0) = v_2$ for all $hv_0 \in \Hist(v_0)\setminus\{v_0\}$.\footnote{As player~$1$ can only loop on vertices $v_2$ and $v_3$, we do not formally define $\sigma_1$ on histories ending with $v_2$ or $v_3$.} Hence player~$1$ chooses to move to $v_1$ (resp. to $v_2$) at the first visit (resp. next visits) to $v_0$. The needed memory is whether the current history has visited $v_0$ once or more time. 
\end{itemize}
The outcome $\out{(\sigma_1,\sigma_2)}_{v_0}$ is equal to $v_0v_1v_3^\omega$ with payoff $p_3$.
\end{example}

\subsection{Studied problems} \label{subsec:probl}

In this paper, we want to study \emph{two problems}. In the first problem, one designated player, say player~$1$, wants to apply a strategy that \emph{guarantees} certain constraints on the payoffs of the plays (with respect to his preference relation) \emph{against} any strategy of the other players. The other players can thus be considered as one player, say player~$2$, being the \emph{opponent} of player~$1$. This is the class of so-called \emph{two-player zero-sum games}.

\begin{problem} \label{prob:twoplayers}
Let $(G,v_0)$ be an initialized two-player zero-sum game and $\mu, \nu \in \setPayoff_1$ be two bounds. Decide whether player~$1$ has a strategy $\sigma_1$ such that $\mu \preceq_1 \fctPayoff_1(\rho)$ (resp. $\mu \preceq_1 \fctPayoff_1(\rho) \preceq_1 \nu$) for all plays $\rho \in \Plays(v_0)$ consistent with $\sigma_1$.\footnote{This problem is focused on Player~$1$, the payoff function $f_2$ and preference relation $\prec_2$ of Player~$2$ do not matter.} 
\end{problem}

Case $\mu \preceq_1 \fctPayoff_1(\rho)$ is called the \emph{threshold problem} whereas case $\mu \preceq_1 \fctPayoff_1(\rho) \preceq_1 \nu$ is called the \emph{constraint problem}. 
When a strategy $\sigma_1$ as required in Problem~\ref{prob:twoplayers} exists, it is called \emph{winning} and a play~$\rho$ consistent with $\sigma_1$  is also called \emph{winning}; we also say that player~$1$ can \emph{ensure} a payoff $\fctPayoff_1(\rho)$ such that $\mu \preceq_1 \fctPayoff_1(\rho)$ (resp. $\mu \preceq_1 \fctPayoff_1(\rho) \preceq_1 \nu$).
When this problem is decidable, we are interested in finding its complexity class and the simplest winning strategies~$\sigma_1$, like positional or finite-memory ones when they exist. 


In a two-player zero-sum game $G$, the opposition between player~$1$ and player~$2$ is most often described in terms of objectives. An \emph{objective}~$\Omega$ for player~$1$ is a subset of $\Plays$, here the set of plays $\rho$ such that $\mu \preceq_1 \fctPayoff_1(\rho)$ (resp. $\mu \preceq_1 \fctPayoff_1(\rho) \preceq_1 \nu$). Player~1 wants to ensure a play in $\Omega$ against any strategy of player~$2$. As an opponent, player~$2$ wants wants to avoid plays in $\Omega$, that is, to ensure the opposite objective $\Plays \setminus \Omega$. We say that the game $G$ with objective $\Omega$ is \emph{determined} if for each initial vertex $v_0$, either player~1 has a winning strategy to ensure $\Omega$ in $(G,v_0)$ or player~2 has a winning strategy to ensure $\Plays \setminus \Omega$. \emph{Martin's theorem}~\cite{Mar75} states that every two-player zero-sum game with \emph{Borel objectives} is determined. Nevertheless, it gives no information on which player has a winning strategy and on the shape of such a winning strategy. This motivates studying Problem~\ref{prob:twoplayers}.


\begin{example} \label{ex:second-ex}
Let us come back to the game of Figure~\ref{fig:gameAbstract} seen as a two-player zero-sum game (we thus focus on player~$1$).  
In $(G,v_0)$, player~$1$ has a winning strategy $\sigma_1$ for the threshold problem with $\mu = p_2$, that is, for the objective $\Omega = \{\rho \mid f(\rho) \in \{p_2,p_3\} \}$: take the positional strategy $\sigma_1$ such that $\sigma_1(v_0) = v_2$. However he has no winning strategy for the threshold problem with $\mu = p_3$. Indeed with the positional strategy $\sigma_2$ such that $\sigma_2(v_1) = v_0$, player~$2$ has a winning strategy for the opposite objective since he can ensure a payoff equal to $p_1$ or $p_2$. 
\end{example}

In the second problem studied in this article, we come back to multi-player games where each player has his own payoff function and preference relation. Here, the players are not necessarily antagonistic: this is the class of so-called \emph{multi-player non zero-sum games}. Instead of looking for a strategy ensuring a certain objective for one designated player, we are now interested in strategy profiles, called \emph{solution profiles}, that provide payoffs satisfactory to all players with respect to their own objectives. A classical example of solution profile is the notion of \emph{Nash equilibrium} (NE)~\cite{Nas50}. Informally, a strategy profile is an NE if no player has an incentive to deviate (with respect to his preference relation) when the other players stick to their own strategies. In other words, an NE can be seen as a \emph{contract} that makes every player satisfied in the sense that nobody wants to break the contract if the others follow it.

\begin{definition} \label{def:NE}
Given an initialized game $(G, v_0)$, a strategy profile $(\sigma_i)_{i \in \Pi}$ is a \emph{Nash equilibrium} if $\fctPayoff_i(\out{(\sigma_i)_{i \in \Pi}}_{v_0}) \nprec_i \fctPayoff_i(\out{\sigma'_i, \sigma_{-i}}_{v_0})$ for all players $i \in \Pi$ and all strategies $\sigma'_i$ of player~$i$.
\end{definition}

In this definition, notation $(\sigma'_i, \sigma_{-i})$ means the strategy profile such that all players stick to their own strategy except player $i$ who shifts from strategy $\sigma_i$ to strategy $\sigma'_i$. We say that $\sigma'_i$ is a \emph{deviating} strategy from $\sigma_i$. When $\fctPayoff_i(\out{(\sigma_i)_{i \in \Pi}}_{v_0}) \prec_i \fctPayoff_i(\out{\sigma'_i, \sigma_{-i}}_{v_0})$, $\sigma'_i$ is called a \emph{profitable deviation} for player~$i$ with respect to $(\sigma_i)_{i \in \Pi}$. 

\addtocounter{example}{-4}
\begin{example}[continued]
Let us reconsider the non zero-sum game $G$ of Figure~\ref{fig:gameAbstract} and the strategy profile $(\sigma_1,\sigma_2)$ given previously in $(G,v_0)$ ($\sigma_1(v_0) = v_1$, $\sigma_1(hv_0) = v_2$ for all $hv_0 \in \Hist(v_0) \setminus \{v_0\}$, and $\sigma_2(hv_1) = v_3$ for all $hv_1 \in \Hist(v_0)$). This strategy profile is an NE with outcome $\out{(\sigma_1,\sigma_2)}_{v_0} = v_0v_1v_3^\omega$. Indeed, player~$1$ has no incentive to deviate since the payoff $p_3$ of $\out{\sigma_1,\sigma_2}_{v_0}$ is the best possible with respect to $\prec_1$. If player~$2$ uses the deviating strategy $\sigma'_2$ from $\sigma_2$ such that $\sigma'_2(v_0v_1) = v_0$, then the resulting outcome $\out{\sigma_1,\sigma'_2}_{v_0} = v_0v_1v_0v_2^\omega$ has a less preferable payoff for him since $p_2 \prec_2 p_3$. So player~$2$ has no profitable deviation.
\end{example}
\addtocounter{example}{+3} 

Other kinds of solution profiles will be studied in Section~\ref{sec:multi}. 

\begin{problem} \label{prob:multiplayers}
Let $(G,v_0)$ be an initialized multi-player non zero-sum game and $(\mu_i)_{i \in \Pi}, (\nu_i)_{i \in \Pi} \in (\setPayoff_i)_{i \in \Pi}$ be two tuples of bounds. Decide whether there exists a solution profile $(\sigma_i)_{i \in \Pi}$ such that $\mu_i \preceq_i \fctPayoff_i(\out{(\sigma_i)_{i \in \Pi}}_{v_0})$ (resp. $\mu_i \preceq_i \fctPayoff_i(\out{(\sigma_i)_{i \in \Pi}}_{v_0}) \preceq_i \nu_i$) for all players $i \in \Pi$. 
\end{problem}

Similarly to Problem~\ref{prob:twoplayers}, the two cases are respectively called \emph{threshold problem} and \emph{constraint problem}, and we want to compute the complexity class and the simplest solution profiles in case of decidability. 


In Sections~\ref{sec:two} and~\ref{sec:multi}, we present some known results about solutions to Problems~\ref{prob:twoplayers} and~\ref{prob:multiplayers} respectively with an emphasis on \emph{general approaches}. Before, we end Section~\ref{sec:prelim} with a list of payoff functions that are classically studied.

\subsection{Classical payoff functions} \label{subsec:obj}


In the classes of games that are \emph{classically} studied, each player $i \in \Pi$ uses a real-valued payoff function $\fctPayoff_i~: \Plays \to \mathbb R$ and a preference relation $\prec_i$ equal to the usual ordering $<$ on $P_i = \mathbb R$. Hence, player~$i$ prefers to maximize the payoff $\fctPayoff_i(\rho)$ of a play $\rho$.\footnote{Alternatively, $\prec_i$ can be the ordering $>$ meaning that player~$i$ prefers to minimize the payoff of a play.} In this classical setting, we focus on two particular subclasses: the \emph{Boolean} payoff functions and the \emph{quantitative} payoff functions.

\subsubsection*{Boolean payoff functions} 

A particular subclass of games $G$ are those equipped with \emph{Boolean} functions $\fctPayoff_i~: \Plays \to \{0,1\}$, for all $i \in \Pi$, where payoff~$1$ (resp. payoff~$0$) means that the play is the most (resp. the less) preferred by player~$i$. Particularly interesting related objectives are $\Omega_i = \{\rho \in \Plays \mid f_i(\rho) = 1 \}$, $i \in \Pi$.
Classical such objectives $\Omega_i$ are \emph{$\omega$-regular objectives} like the following ones~\cite{2001automata,GU08,PP04}. 

\begin{definition} \label{def:omega-reg}
\begin{itemize}
\item Let $U \subseteq V$,
\begin{itemize}
\item \emph{Reachability}~: $\Omega_i = \{\rho \in \Plays \mid \rho$ visits a vertex of $U$ at least once$\}$,
\item \emph{Safety}: $\Omega_i = \{\rho \in \Plays \mid \rho$ visits no vertex of $U \}$,
\item \emph{B\"uchi}: $\Omega_i = \{\rho \in \Plays \mid \Infrho(\rho) \cap U \neq \emptyset \}$,
\item \emph{Co-B\"uchi}: $\Omega_i = \{\rho \in \Plays \mid \Infrho(\rho) \cap U = \emptyset \}$.
\end{itemize}
\item Let $\fctCol~: V \rightarrow \mathbb N$ be a coloring of the vertices by integers,
\begin{itemize}
\item  \emph{Parity}: $\Omega_i = \{\rho \in \Plays \mid$ the maximum color seen infinitely often along $c(\rho_0)c(\rho_1) \ldots $ is even$\}$.
\end{itemize}
\item Let $(F_k, G_k)_{1 \leq k \leq l}$ be a family of pairs of sets $F_k, G_k \subseteq V$,
\begin{itemize}
\item \emph{Rabin}: $\Omega_i = \{\rho \in \Plays \mid \exists k$, $1 \leq k \leq l$, such that  $\Infrho(\rho) \cap F_k = \emptyset$ and $\Infrho(\rho) \cap G_k \neq \emptyset \}$,
\item \emph{Streett}: $\Omega_i = \{\rho \in \Plays \mid \forall k$, $1 \leq k \leq l$, $\Infrho(\rho) \cap F_k \neq \emptyset$ or $\Infrho(\rho) \cap G_k = \emptyset \}$.
\end{itemize}
\item Let ${\cal F} \subseteq 2^V$ be a family of subsets of vertices,
\begin{itemize}
\item \emph{Muller}\footnote{A \emph{colored variant} of Muller objective is defined from a coloring $\fctCol~: V \rightarrow \mathbb N$ of the vertices: the family $\cal F$ is composed of subsets of $\fctCol(V)$ (instead of $V$) and $\Omega_i = \{\rho \in \Plays \mid \Infrho(c(\rho_0)c(\rho_1)\ldots) \in {\cal F} \}$~\cite{GU08}. See~\cite{HunterD05} for several variants of Muller games.}%
: $\Omega_i = \{\rho \in \Plays \mid \Infrho(\rho) \in {\cal F} \}$.
\end{itemize}
\end{itemize}
\end{definition}

Notice that reachability and safety (resp. B\"uchi and co-B\"uchi, Rabin and Streett) are dual objectives. The complement of a parity (resp. Muller) objective is again a parity (resp. Muller) objective: from the coloring function $\fctCol~: V \rightarrow \mathbb N$, define the new function $\fctCol'$ such that $\fctCol'(v) = \fctCol(v)+1$ for all $v \in V$ (resp. from the family  ${\cal F} \subseteq 2^V$, define the new family ${\cal F'} = 2^V \setminus {\cal F}$). A B\"uchi (resp. co-B\"uchi) objective is a particular case of a parity objective: assign color~$2$ to vertices of $U$ and~$1$ to vertices of $V \setminus U$ (resp. color~$1$ to $U$ and~$0$ to $V \setminus U$). Similarly, one can easily prove that a parity objective is both a Rabin and a Streett objective which are themselves a Muller objective~\cite{2001automata}. 

In the previous definition, the payoff function $\fctPayoff_i$ is prefix-independent in each case except for reachability and safety where only condition (\ref{eq:1}) of prefix-linearity is satisfied.

\begin{example} \label{ex:third-ex}
Suppose that in the game of Figure~\ref{fig:gameAbstract}, player~$1$ wants to achieve the B\"uchi objective with $U = \{v_2,v_3\}$ whereas player~$2$ wants to achieve the Muller objective with ${\cal F} = \{ \{v_0,v_1\},\{v_3\}\}$. Then the play $\rho = (v_0v_1)^\omega$ has payoff $(0,1)$, that is a payoff 0 for player~$1$ and a payoff~$1$ for player~$2$.
\end{example}

\subsubsection*{Quantitative payoff functions}

Classical \emph{quantitative} payoff functions $\fctPayoff_i : \Plays \to \mathbb R$ are defined from a \emph{weight function} $w_i~: E \to \mathbb Q$ as follows~\cite{LaurentDoyen} (each edge of the game $G$ is thus labeled by a $|\Pi|$-tuple of weights). 

\begin{definition} \label{def:functions}
Let $w_i~: E \to \mathbb Q$ be a weight function and $\lambda \in \; ]0,1[$ be a rational \emph{discount factor}. Then $\fctPayoff_i~: \Plays \to \mathbb R$ is defined as one among the following payoff functions: let $\rho = \rho_0\rho_1\ldots \in \Plays$,
\begin{itemize} 
\item \emph{Supremum}: $\Sup_i(\rho) =  \sup_{n \in \mathbb N} w_i(\rho_n, \rho_{n+1})$,
\item \emph{Infimum}: $\Inf_i(\rho)  = \inf_{n \in \mathbb N} w_i(\rho_n, \rho_{n+1})$,
\item \emph{Limsup}: $\LimSup_i(\rho)  =  \limsup\limits_{n \to \infty} w_i(\rho_n, \rho_{n+1})$, 
\item \emph{Liminf}: $\LimInf_i(\rho) = \liminf\limits_{n \to \infty} w_i(\rho_n, \rho_{n+1})$,
\item \emph{Mean-payoff $\SupMP_i$}: $\SupMP_i(\rho)  =  \limAvgSup w_i(\rho_k, \rho_{k+1})$,
\item \emph{Mean-payoff $\InfMP_i$}: $\InfMP_i(\rho)  =  \limAvgInf w_i(\rho_k, \rho_{k+1})$,

\item \emph{Discounted sum}: $\Disc^\lambda_i(\rho) =  \sum_{n=0}^{\infty} w_i(\rho_n, \rho_{n+1}) \lambda^n $.  
\end{itemize}
\end{definition}

Some of these payoff functions provide natural generalizations of the previous $\omega$-regular objectives. Indeed the supremum (resp. infimum, limsup, liminf) function is a quantitative generalization of the reachability (resp. safety, B\"uchi, co-B\"uchi) objective. The mean-payoff and discounted sum functions are much studied in classical game theory~\cite{filar1997}. 

There are two variants of mean-payoff functions because the limit may not exist. Nevertheless in case of a lasso $\rho = hg^\omega$, both payoffs $\SupMP_i(\rho)$ and $\InfMP_i(\rho)$ coincide and are equal to the average weight of the cycle $g$ (with respect to the weight function $w_i$). 

In Definition~\ref{def:functions}, the payoff function $\fctPayoff_i$ is prefix-independent in limsup, liminf and mean-payoff cases, prefix-linear in discounted sum case, and satisfies condition (\ref{eq:1}) of prefix-linearity in supremum and infimum cases.

\begin{example} \label{ex:4-ex}
We equip the game of Figure~\ref{fig:gameAbstract} with two weight functions $w_1, w_2$, leading to the game of Figure~\ref{fig:gameWeight}. Suppose that $f_1 = \LimSup_1$ and $f_2= \SupMP_2$. The preferences of the players with respect to plays $(v_0v_1)^\omega$ and $v_0v_2^\omega$ are opposed since $\fctPayoff_1((v_0v_1)^\omega) = 1 < \fctPayoff_1(v_0v_2^\omega) = 2$ for player~$1$, and $\fctPayoff_2(v_0v_2^\omega) = 1 < \fctPayoff_2((v_0v_1)^\omega) = 3$  for player~$2$.

\begin{figure}[ht!]
\begin{center}
\begin{tikzpicture}[initial text=,auto, node distance=2cm, shorten >=1pt]

\node[state, scale=0.6]              (1)                     {$v_0$};
\node[state, rectangle, scale=0.6]   (2)    [right=of 1]     {$v_1$};
\node[state, scale=0.6]              (3)    [left=of 1]      {$v_2$};
\node[state, scale=0.6]              (4)    [right=of 2]     {$v_3$};

\path[->] (1) edge [bend right=25]          node[below, scale=0.7, black]    {$(1, 3)$}   (2)
             edge                                node[above, scale=0.7]           {$(0,0)$}   (3)

         (2) edge  [bend left=-25]               node[above, scale=0.7]           {$(1, 3)$}   (1)
             edge  []                        node[above, scale=0.7, black]    {$(0,0)$}   (4)

         (3) edge  [loop above]            node[midway, scale=0.7, black]   {$(2,1)$}   ()

         (4) edge  [loop above]            node[midway, scale=0.7, black]   {$(3,2)$}   ();

\end{tikzpicture}
\caption{A quantitative two-player game}
\label{fig:gameWeight}
\end{center}
\end{figure}
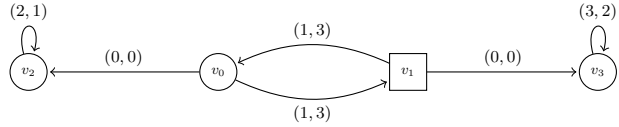

\end{example}

In the sequel, games with the Boolean payoff functions of Definition~\ref{def:omega-reg} are called \emph{Boolean games}. Similarly games with the quantitative payoff functions of Definition~\ref{def:functions} are called \emph{quantitative games}. We also speak about \emph{reachability game}, \emph{supremum game}, etc, when we want to refer to a game where \emph{all} the players use the \emph{same} type of payoff function. The complexity results mentioned later depend on the number of vertices, edges and players, as well as on the number of colors (resp. pairs, elements of $\cal F$) for parity (resp. Rabin/Streett, Muller) games, and on numerical rational values (of weights, discount factor, and bounds) given in binary for quantitative games.

\section{Two-player zero-sum games} \label{sec:two}

In two-player zero-sum games, players~$1$ and~$2$ have opposite objectives. This class of games has been much studied. In particular solutions to Problem~\ref{prob:twoplayers} are well established for Boolean games and quantitative games as introduced in Section~\ref{subsec:obj}. Before presenting them, we begin with the simplest situation of games played by a \emph{unique player} and we show that the problems studied in this article are connected to problems in automata theory and numeration systems. 

\subsection{One-player games} \label{subsec:one}

In \emph{one-player games}, player~1 has no opponent, he is the only player to choose the next vertex at any moment of a play. In other words, a strategy $\sigma_1$ for player~$1$ is nothing else than a play $\rho$ in the game. The statement of Problem~\ref{prob:twoplayers} thus simplifies as follows:\footnote{In Section~\ref{subsec:one}, we omit index~$1$ everywhere since player~$1$ is the unique player of the game.}

\begin{problem} \label{prob:oneplayer}
Let $(G,v_0)$ be an initialized one-player game. Let $\mu, \nu \in \setPayoff$ be two bounds.
Decide whether there exists a play $\rho \in \Plays(v_0)$ such that $\mu \preceq \fctPayoff(\rho)$ (resp. $\mu \preceq \fctPayoff(\rho) \preceq \nu$)?
\end{problem}


\subsubsection{Boolean games} For Boolean games, this problem is interesting only with bounds $\mu = \nu = 1$. Indeed recall that the payoff function $f$ is Boolean and that player~$1$ prefers plays $\rho$ such that $f(\rho) = 1$. This is the classical well-known \emph{non emptiness problem for automata}~\cite{PP04}. For instance, Problem~\ref{prob:oneplayer} for one-player reachability (resp. B\"uchi) games with $\mu = \nu = 1$ is the non emptiness problem for automata accepting finite words (resp. B\"uchi automata accepting infinite words).

\begin{theorem} \label{thm:onePlayerOmegaReg}
Let $(G,v_0)$ be an initialized one-player Boolean game. Then Problem~\ref{prob:oneplayer} (with $\mu = \nu = 1$) is decidable in polynomial time with positional winning strategies, except for Streett and Muller games where finite-memory strategies are necessary and sufficient.
\end{theorem}

Let us comment this theorem. Notice that a winning strategy for player~1 that is finite-memory (resp. positional) means that the corresponding winning play $\rho$, or in terms of automata the accepted word, is a (resp. simple) lasso. It is well-known that positional strategies are sufficient for B\"uchi objectives. This also happens for the other objectives except for Streett and Muller objectives (we will discuss this point in more details in Section~\ref{subsec:two}, see Theorem~\ref{thm:TwoPlayerOmegaReg}). Example~\ref{ex:mem} illustrates that finite-memory strategies are necessary for Streett and Muller games. In cases where positional strategies are sufficient, an algorithm for Problem~\ref{prob:oneplayer} has thus to concentrate on the existence of winning simple lassos, which can be easily done in polynomial time. The case of Streett and Muller games can also be solved in polynomial time~\cite{EmersonL87,Horn08}. Problem~\ref{prob:oneplayer} is {\sf NL}-complete for reachability and B\"uchi games~\cite{Jones75,VardiW94} as well as for safety, co-B\"uchi, Rabin, parity, and Muller games, and it is {\sf P}-complete for Streett games~\cite{EmersonL87,SafraV89}.


\begin{example} \label{ex:mem} Consider the initialized one-player game $(G,v_0)$ of Figure~\ref{fig:gameOneplayer} with $V = \{v_0,v_1,v_2\}$. For the Muller objective with ${\cal F} = \{V\}$  (or the Streett objective with the two pairs $(F_1,G_1), (F_2,G_2)$ such that $F_1= \{v_1\}$, $G_1 =V$ and $F_2 = \{v_2\}$, $G_2 = V$), a winning play $\rho \in \Plays(v_0)$ cannot be a simple lasso as it has to alternate between $v_1$ and $v_2$.

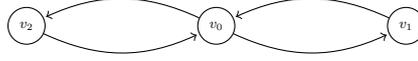
\begin{figure}
\begin{center}
\begin{tikzpicture}[initial text=,auto, node distance=2cm, shorten >=1pt]

\node[state, scale=0.6]              (1)                     {$v_0$};
\node[state, scale=0.6]              (2)    [right=of 1]     {$v_1$};
\node[state, scale=0.6]              (3)    [left=of 1]      {$v_2$};

\path[->] (1) edge [bend right=25]          node[below, scale=0.7, black]    {}   (2)
             edge [bend right=25]                     node[above, scale=0.7]           {}   (3)

         (2) edge  [bend left=-25]               node[above, scale=0.7]           {}   (1)

         (3) edge  [bend left=-25]            node[midway, scale=0.7, black]   {}   (1);
\end{tikzpicture}
\caption{A one-player game}
\label{fig:gameOneplayer}
\end{center}
\end{figure}
\end{example}    

%

%
%
%
 
\subsubsection{Quantitative games} Let us turn to quantitative games. The existence of plays $\rho$ with $\mu \leq \fctPayoff(\rho)$ in one-player quantitative games (threshold problem) have been studied in~\cite{LaurentDoyen}. 

\begin{theorem} \label{thm:LaurentDoyen}  \cite{LaurentDoyen}
Let $(G,v_0)$ be an initialized one-player quantitative game, and $\mu \in \mathbb Q$ be a rational threshold. Then deciding whether there exists a play $\rho \in \Plays(v_0)$ such that $\mu \leq \fctPayoff(\rho)$ is solvable in polynomial time with positional strategies.
\end{theorem}

Let us comment this theorem. Dealing with functions $\Sup$, $\Inf$, $\LimSup$, and $\LimInf$ is equivalent to respectively consider reachability, safety, B\"uchi, and co-B\"uchi objectives (studied in Theorem~\ref{thm:onePlayerOmegaReg}). For instance, satisfying $\mu \leq \Sup(\rho)$ is equivalent to visiting an edge with a weight $\geq \mu$ along $\rho$. For functions $\SupMP$, $\InfMP$, and $\Disc^\lambda$, once one knows that positional strategies are sufficient (we will discuss this point in more details in Section~\ref{subsec:two}), the problem again reduces to the existence of a simple lasso $\rho = hg^\omega$ with maximum payoff $\fctPayoff(\rho)$. In case of mean-payoff function, recall that both payoffs $\SupMP(\rho)$, $\InfMP(\rho)$ coincide and are equal to the average weight of the cycle $g$. A polynomial algorithm is proposed in~\cite{karp} to compute a cycle in a weighted graph with maximum average weight. The case of function $\Disc^\lambda$ is polynomially solved by a linear programming approach in~\cite{And06}.

We now discuss the existence of a play $\rho$ such that $\mu \leq \fctPayoff(\rho) \leq \nu$, given two rational bounds $\mu, \nu \in \mathbb Q$ (constraint problem). The problem is more involved, in particular it is currently unsolved for function $\Disc^\lambda$.

\begin{theorem} \cite{HunterR14,UmmelsArXiv}
Let $(G,v_0)$ be an initialized one-player quantitative (except discounted sum) game, and $\mu, \nu \in \mathbb Q$ be two rational bounds. Then deciding whether there exists a play $\rho \in \Plays(v_0)$ such that $\mu \leq \fctPayoff(\rho) \leq \nu$ is solvable in polynomial time. Positional strategies are sufficient for supremum, infimum, limsup, and liminf games, whereas finite-memory is necessary and sufficient for mean-payoff $\SupMP$ and $\InfMP$ games.
\end{theorem}

Let us comment this theorem. If we focus on function $\LimSup$, looking for a play $\rho$ such that $\mu \leq \LimSup(\rho) \leq \nu$ reduces to the non emptiness problem for Rabin automata (studied in Theorem~\ref{thm:onePlayerOmegaReg}). Indeed the required play $\rho$ is such that at least one weight seen infinitely often along $\rho$ is $\geq \mu$ and none of them is $> \nu$. A similar approach exists for functions $\Sup$, $\Inf$ and $\LimInf$. Whereas positional winning strategies are sufficient in all these cases, finite-memory is needed for mean-payoff functions as indicated in Example~\ref{ex:memMP}.  Since finite-memory strategies are sufficient~\cite{HunterR14}, the problem in both cases $\SupMP$, $\InfMP$ reduces to the existence of a lasso $\rho$ satisfying the constraints. This can be checked in polynomial time by solving a linear program~\cite{UmmelsArXiv}.   

\begin{example} \label{ex:memMP} Consider the game of Figure~\ref{fig:gameOneplayer} equipped with the weight function $w$ that labels the two left edges by $0$ and the two right edges by $2$. A winning play~$\rho$ for $\mu = \nu = 1$ cannot be a simple lasso (with payoff either $0$ or $2$). However the non simple lasso $\rho = (v_0v_1v_0v_2)^\omega$ is winning.
\end{example}    

Concerning function $\Disc^\lambda$, Problem~\ref{prob:oneplayer} is open. It is closely related to the following open problem, called \emph{target discounted-sum problem} in~\cite{BokerHO15}.

\begin{problem} \label{prob:Boker}
Given three rational numbers $a,b$ and $t$, and a rational discount factor $\lambda \in \, ]0,1[$, does there exist an infinite sequence $u = u_0u_1 \ldots \in \{a, b\}^\omega$ such that $\sum_{n=0}^{\infty}  u_n \lambda^n$ is equal to $t$?
\end{problem}

The authors of~\cite{BokerHO15} show that Problem~\ref{prob:Boker} is related to several open questions in mathematics and computer science. In particular it is related to \emph{numeration systems} and more precisely to $\beta$-representations of real numbers~\cite{BR16,Lothaire}. Given $\beta > 1$ a real number (the base) and $A \subseteq \mathbb N$ a finite alphabet (the set of digits), a \emph{$\beta$-representation} of a real number $x \geq 0$ is an infinite sequence $(x_n)_{n \leq k} \in A^\omega$, also written $x_k \ldots x_0 . x_{-1}x_{-2} \ldots$,  such that $x = \sum_{n \leq k} x_n \beta^n$.  A well-known  result~\cite{Renyi57} is that every $x \geq 0$ has a $\beta$-representation using $A = \{0,1,\ldots, \lceil \beta - 1 \rceil \}$. It follows that Problem~\ref{prob:Boker} asks whether $t$ has a $\beta$-representation $x_0 . x_{-1}x_{-2} \ldots$ (with $k=0$) using $\beta = \frac{1}{\lambda}$ and $A = \{a,b\}$. This problem is therefore decidable when $a = 0, b = 1$ and $\lambda \geq \frac{1}{2}$. Indeed using the result of~\cite{Renyi57}, either $t > \frac{1}{\beta - 1}$ and it has no $\beta$-representation $x_0 . x_{-1}x_{-2} \ldots \in \{0,1\}^\omega$, or $t \leq \frac{1}{\beta - 1}$ and it has such a $\beta$-representation. Other partial results to Problem~\ref{prob:Boker} can be found in~\cite{BokerHO15}.

\subsection{Two-player games} \label{subsec:two}

We now turn to two-player zero-sum games. In Problem~\ref{prob:twoplayers}, the objective of player~1 is the set $\Omega$ of plays $\rho$ such that $\mu \preceq_1 \fctPayoff_1(\rho)$ (resp. $\mu \preceq_1 \fctPayoff_1(\rho) \preceq_1 \nu$), whereas player~2 has the opposite objective $\Plays \setminus \Omega$. Examples of the threshold problem are the following ones: in a reachability game, player~1 aims at reaching some target set of vertices whereas player~2 tries to prevent him from reaching it; in a limsup game, player~1 aims at maximize the payoff $\LimSup(\rho)$ of the play $\rho$ (in a way to be $\geq \mu$) whereas player~2 tries to minimize it. Recall that by Martin's theorem, every two-player zero-sum games with Borel objectives is determined. This large class of games includes the objectives $\Omega$ of player~$1$ in Problem~\ref{prob:twoplayers} for the Boolean and quantitative games introduced in Section~\ref{subsec:obj}.  A lot of research has been developed to solve Problem~\ref{prob:twoplayers} that we present in this section. 
In Sections~\ref{subsec:two} and~\ref{subsec:preorders}, as the objectives $\Omega$ and $\Plays \setminus \Omega$ of players~$1$ and~$2$ only depend on $f_1$, $\prec_1$, and $P_1$, we simplify the used notation by omitting index~$1$. 

\subsubsection{Criterium for uniform optimal strategies}
We begin by studying the winning strategies that player~$1$ can use for the threshold problem in Problem~\ref{prob:twoplayers}. This is related to the notion of \emph{value} and \emph{optimal} strategy.

\begin{definition} \label{def:value}
Let $(G,v_0)$ be an initialized two-player zero-sum game. If there exists $\val(v_0) \in \setPayoff$ such that
\begin{itemize}
\item player~$1$ has a strategy $\sigma_1$ such that $\val(v_0) \preceq \fctPayoff(\rho)$ for all plays $\rho$ in $\Plays(v_0)$ consistent with $\sigma_1$, and
\item player~$2$ has a strategy $\sigma_2$ such that $\fctPayoff(\rho) \preceq \val(v_0)$ for all plays $\rho$ in $\Plays(v_0)$ consistent with $\sigma_2$,
\end{itemize}
then $\val(v_0) = \fctPayoff(\out{\sigma_1,\sigma_2}_{v_0})$ is the \emph{value} of $v_0$ and $\sigma_1$ (resp. $\sigma_2$) is an \emph{optimal strategy} for player~$1$ (resp. player~$2$). 
\end{definition}

Intuitively, $\val(v_0)$ is the highest threshold $\mu$ for which player~$1$ can ensure (with an optimal strategy) a payoff $\fctPayoff(\rho)$ such that $\mu \preceq \fctPayoff(\rho)$. In this definition, the antagonistic player~$2$ behaves in the opposite way. When the value $\val(v_0)$ exists and is computable, the threshold problem is easily solved: we just check whether the given threshold $\mu$ satisfies $\mu \preceq \val(v_0)$. Moreover both players can limit themselves to use optimal strategies, that is, if player~$1$ has a winning strategy (resp. no winning strategy) for the threshold problem, then player~$1$ (resp. player~$2$) can use an optimal strategy as winning strategy (resp. for the opposite objective). 

\addtocounter{example}{-15} 
\begin{example}[continued]  \label{ex:values}
Let us come back to the two-player zero-sum game of Example~\ref{ex:second-ex}. Recall that in $(G,v_0)$, player~$1$ has a winning strategy for the threshold problem with $\mu = p_2$ but not with $\mu = p_3$, meaning that $\val(v_0) = p_2$. Indeed, one can check that $\val(v_0) = \val(v_1) = \val(v_2) = p_2$ and $\val(v_3) = p_3$, and that both players have optimal strategies that are positional, and even more uniform. The values are indicated under the vertices in Figure~\ref{fig:gameValue}, and the two uniform optimal strategies are given as thick edges.
\begin{figure}[ht!]
\begin{center}
\begin{tikzpicture}[initial text=,auto, node distance=2cm, shorten >=1pt]

\node[state, scale=0.6]              (1)                     {$v_0$}; 
\node[state, rectangle, scale=0.6]   (2)    [right=of 1]     {$v_1$};
\node[state, scale=0.6]              (3)    [left=of 1]      {$v_2$};
\node[state, scale=0.6]              (4)    [right=of 2]     {$v_3$};
\node[scale=0.7]                                  (8)     [below=.1cm of 1] {$p_2$}; 
\node[scale=0.7]                                  (8)     [below=.1cm of 2] {$p_2$}; 
\node[scale=0.7]                                  (8)     [below=.1cm of 3] {$p_2$}; 
\node[scale=0.7]                                  (8)     [below=.1cm of 4] {$p_3$}; 

\path[->] (1) edge [bend right=25]          node[below, scale=0.7, black]    {}   (2)
             edge  [thick, black]                                  node[above, scale=0.7]           {}   (3)

         (2) edge  [bend left=-25,thick, black]               node[above, scale=0.7]           {}   (1)
             edge                                                  node[above, scale=0.7, black]    {}   (4)

         (3) edge  [loop above,thick, black]            node[midway, scale=0.7, black]   {}   ()

         (4) edge  [loop above, thick, black]            node[midway, scale=0.7, black]   {}   ();

\end{tikzpicture}
\caption{Values and uniform optimal strategies for a two-player zero-sum game with $\fctPayoff((v_0v_1)^\omega) = p_1\prec\fctPayoff(v_2^\omega) = p_2 \prec \fctPayoff(v_3^\omega) = p_3$}
\label{fig:gameValue}
\end{center}
\end{figure}
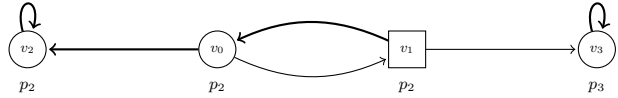
\end{example}
\addtocounter{example}{+14} 

We will see later in this section that Boolean and quantitative games often have uniform optimal strategies (see Theorems~\ref{thm:TwoPlayerOmegaReg} and~\ref{thm:twoquantitative}). In~\cite{GimbertZ04}, the authors propose a \emph{unified approach} to all these results: they give a general criterium on the payoff function that guarantees uniform optimal strategies for both players.

\begin{theorem}\cite{GimbertZ04}\footnote{The hypotheses of this theorem are those given in the full version of \cite{GimbertZ04} available at http://www.labri.fr/perso/gimbert/} \label{thm:mixing}
Let $G$ be a two-player zero-sum game with a preference relation $\prec$ on $P$ such that each subset of $\setPayoff$ has an infimum and a supremum. If the payoff function $f$ is \emph{fairly mixing}, that is,
\begin{enumerate}
\item $\forall h\rho, h\rho' \in \Plays$, if $\fctPayoff(\rho) \preceq \fctPayoff(\rho')$ then $\fctPayoff(h\rho)  \preceq \fctPayoff(h \rho')$,
\item $\forall h\rho, h\rho' \in \Plays$, $\min\{ \fctPayoff(\rho), \fctPayoff(h^\omega) \}  \preceq \fctPayoff(h \rho)  \preceq \max\{ \fctPayoff(\rho), \fctPayoff(h^\omega) \}$,
\item $\forall h_k \in \Hist, k \in \mathbb N$, 

$\min\{ \fctPayoff(h_0h_2h_4 \ldots), \fctPayoff(h_1h_3h_5 \ldots), \inf_k \fctPayoff(h_k^\omega) \}$ 

$\quad  \preceq \fctPayoff(h_0h_1h_2h_3 \ldots)$ 

$\quad \preceq \max\{ \fctPayoff(h_0h_2h_4 \ldots), \fctPayoff(h_1h_3h_5 \ldots), \sup_k \fctPayoff(h_k^\omega) \}$,
\end{enumerate}
then both players have uniform optimal strategies.
\end{theorem}

Let us comment this theorem. The first condition is condition (\ref{eq:1}) of prefix-linearity. If $\fctPayoff$ is prefix-independent, then the first and the second conditions are trivially satisfied. The third condition is concerned with shuffles of histories. Let us apply this theorem to quantitative games, for instance to function $\LimSup$ (see Definition~\ref{def:functions}). 
This function is prefix-independent and satisfies the third condition since $\inf_k \LimSup(h_k^\omega) \leq \LimSup(h_0h_1h_2h_3 \ldots) \leq \sup_k \LimSup(h_k^\omega)$. One can check that the payoff functions of all quantitative games are fairly mixing, as well as the payoff functions of the Boolean games with reachability, safety, B\"uchi, co-B\"uchi, and parity objectives~\cite{GimbertZ04} (but not with Streett and Muller objectives).

The proof\footnote{Theorem~\ref{thm:mixing} is given in \cite{GimbertZ04} for real-valued payoff functions $f : \Plays \to \mathbb R$ and the usual ordering $<$, but its proof is easily generalized to the statement given here.} of Theorem~\ref{thm:mixing} is simple and elegant; it is by induction on $|E|-|V|$. If $|E|=|V|$ then there is exactly one outgoing edge for each vertex and thus both players have a unique possible strategy that is therefore uniform and optimal. Suppose that $|E| > |V|$ and let us focus on player~$1$ (a symmetric argument is used for player~$2$). If all vertices $v \in V_1$ have only one outgoing edge, then player~$1$ has a unique strategy, and it is uniform and optimal. Suppose that some $v \in V_1$ has at least two outgoing edges. We partition this set of edges into two non empty subsets $E'_v$ and $E''_v$. From $G$ we define two smaller games $G'$ and $G''$ with the same vertices and edges except that the set of outgoing edges from $v$ is restricted to $E'_v$ in $G'$ and to $E''_v$ in $G''$. By induction hypothesis, $v$ has a value $\val'(v)$ in $G'$ and $\val''(v)$ in $G''$, and both players have uniform optimal strategies, respectively $\sigma'_1, \sigma'_2$ in $G'$ and $\sigma''_1, \sigma''_2$ in $G''$. W.l.o.g. $\val''(v) \preceq \val'(v)$, we then choose $\sigma'_1$ as optimal strategy for player~$1$ in $G$ and for all $u \in V$, we take their value $\val'(u)$ in $G'$ as their value in $G$. Clearly $\sigma'_1$ is optimal and uniform in $G$. The rest of the proof consists in defining a strategy for player~$2$ (from $\sigma'_2$ and $\sigma''_2$) that is optimal in $G$. This is possible thanks to the three conditions of Theorem~\ref{thm:mixing} applied on plays decomposed according to occurrences of $v$.

Further results can be found in \cite{GimbertZ05}: a characterization of payoff functions is given guaranteeing the existence of uniform optimal strategies for both players. From this characterization, it follows that if both players have uniform optimal strategies when playing solitary in one-player games, then they also have uniform optimal strategies in zero-sum two-player games. 


\subsubsection{Boolean games}
Let us now focus on Boolean games. As for one-player games, we limit the study of Problem~\ref{prob:twoplayers} (threshold and constraint problems) to the only interesting case $\mu = \nu = 1$. The following theorem for two-player games is the counterpart of Theorem~\ref{thm:onePlayerOmegaReg} for one-player games. 

\begin{theorem} \label{thm:TwoPlayerOmegaReg}
Let $(G,v_0)$ be an initialized two-player zero-sum Boolean game. Then Problem~\ref{prob:twoplayers} (with $\mu = \nu = 1$) is
\begin{itemize}
\item {\sf P}-complete with uniform winning strategies for reachability, safety, B\"uchi, and co-B\"uchi objectives~\cite{Beeri,EJ91,2001automata,Immerman81},
\item {\sf P}-complete with finite-memory winning strategies for Muller\footnote{It is {\sf PSPACE}-complete for the colored variant of Muller objective~\cite{HunterD05,McNaughton93}.} objective~\cite{Horn08},
\item {\sf NP}-complete with uniform winning strategies for Rabin objective~\cite{Emerson85,EJ91},
\item {\sf co-NP}-complete with finite-memory winning strategies for Streett objective~\cite{BuhrkeLV96,EJ91},
\item in {\sf NP $\cap$ co-NP} with uniform winning strategies for parity objectives~\cite{EJ91}.
\end{itemize}
\end{theorem}

Let us comment this theorem. The existence of uniform winning strategies (for all objectives except Rabin and Muller objectives) was previously mentioned as a consequence of Theorem~\ref{thm:mixing}~\cite{GimbertZ04}.
Notice that here a value $\val(v_0) = 1$ is equivalent to say that  player~$1$ has a winning strategy for Problem~\ref{prob:twoplayers}. In case player~$1$ has no winning strategy ($\val(v_0) = 0$), it follows that player~$2$ has a winning strategy for the opposite objective by Martin's theorem. Hence Theorem~\ref{thm:TwoPlayerOmegaReg} also gives information for player~$2$ by considering the opposite objective. 
In~\cite{Kopczynski06}, the author gives general conditions on Boolean objectives that guarantee the existence of a uniform winning strategy for one of the players (and not necessarily for both players). This includes the case of Rabin games where player~$1$ has a uniform winning strategy (whereas player~$2$ needs to use a finite-memory strategy to win the opposite Streett objective).

Problem~\ref{prob:twoplayers} is decidable in $O(|V|+|E|)$ time for reachability and safety games~\cite{2001automata}, and the current best algorithm for B\"uchi and co-B\"uchi games is in $O(|V|^2)$ time~\cite{ChatterjeeH14}. For Muller games with ${\cal F} \subseteq 2^V$, the complexity is in $O(|{\cal F}|\cdot(|{\cal F}| + |V| \cdot |E|)^2)$ time~\cite{Horn08}, whereas for Rabin and Streett games with $l$ pairs $(F_k,G_k)$, it is in $O(|V|^{l+1}l!)$ time~\cite{PitermanP06}. Concerning parity games, the complexity class of Problem~\ref{prob:twoplayers} is refined to {\sf UP $\cap$ co-UP} in~\cite{Jurdzinski98} and a major open problem is whether it can be solved in polynomial time. Very recently, a breakthrough quasi-polynomial time algorithm has been proposed in~\cite{Calude17} for parity games. 

\subsubsection{Quantitative games}
Let us turn to quantitative games for which we first give results for the threshold problem, and then for the constraint problem. The following theorem provides the known results to the threshold problem. It describes the simplest form of winning strategies for player~$1$ (resp. player~$2$) when he has a winning strategy for this problem (resp. for ensuring the opposite objective when player~$1$ has no winning strategy).

\begin{theorem} \label{thm:twoquantitative}
Let $(G,v_0)$ be an initialized two-player zero-sum quantitative game, and $\mu \in \mathbb Q$ be a rational bound. Then the threshold problem (in Problem~\ref{prob:twoplayers})~is 
\begin{itemize}
\item {\sf P}-complete for supremum, infimum, limsup, and liminf games with uniform winning strategies for both players,
\item in {\sf NP $\cap$ co-NP} for mean-payoff and discounted sum games with uniform winning strategies for both players~\cite{ZP96}.
\end{itemize}
\end{theorem}

Let us comment this theorem. 
We already know the existence of uniform winning strategies from Theorem~\ref{thm:mixing}~\cite{GimbertZ04}. The {\sf P}-completeness for supremum, infimum, limsup, and liminf games follows from the {\sf P}-completeness for reachability, safety, B\"uchi, and co-B\"uchi games in Theorem~\ref{thm:TwoPlayerOmegaReg}. 
Parity games  are polynomially reducible to mean-payoff games~\cite{Jurdzinski98} which are themselves polynomially reducible to discounted sum games~\cite{ZP96}. For these three classes of games, from the existence of uniform winning strategies, we get a threshold problem in {\sf NP} as follows: guess a uniform strategy $\sigma_1$ for player~$1$ (by choosing one outgoing edge $(v,v')$ for all $v \in V_1$), fix this strategy $\sigma_1$ in the game $G$ to get a one-player game $G_{\sigma_1}$, apply the related polynomial time algorithm of Theorems~\ref{thm:onePlayerOmegaReg} or~\ref{thm:LaurentDoyen} (from the point of view of player~$2$ who controls $G_{\sigma_1}$). The {\sf co-NP} membership is symmetrically obtained with player~$2$.

Concerning the constraint problem, recall that it is more complex already for one-player games (see Section~\ref{subsec:one}) with no known solution for discounted sum games (see Problem~\ref{prob:Boker}). 

\begin{theorem} \label{thm:twoquantitativeconstraint}
Let $(G,v_0)$ be an initialized two-player zero-sum quantitative (except discounted sum) game, and $\mu,\nu \in \mathbb Q$ be rational bounds. Then the constraint problem (in Problem~\ref{prob:twoplayers}) is 
\begin{itemize}
\item {\sf P}-complete for supremum, infimum, limsup, and liminf games with uniform winning strategies for both players~\cite{BruyereHR16,HunterR14},
\item in {\sf NP $\cap$ co-NP} for mean-payoff games with finite-memory (resp. uniform) winning strategies for player~$1$ (resp. player~$2$) \cite{HunterR14}.
\end{itemize}
\end{theorem}

Discounted sum games are studied with bounds $\mu, \nu$ such that $\mu < \nu$ (to avoid the case $\mu = \nu$ of Problem~\ref{prob:Boker}) in~\cite{HunterR14} where it is proved that the constraint problem is {\sf PSPACE}-complete with finite-memory winning strategies for both players.

\subsection{Variants of preferences} \label{subsec:preorders}

Several extensions\footnote{The reader who prefers to know classical solutions to Problem~\ref{prob:multiplayers} for multi-player non zero-sum games can skip this section and go directly to Section~\ref{sec:multi}.} of two-player zero-sum Boolean and quantitative games have been studied in the literature, by using preferences that are irreflexive and transitive but \emph{not necessarily total}, or more generally by using \emph{preorders} $\preceq$ that are reflexive and transitive binary relations (hence, $\preceq$ is not supposed to be total and one can have $p \preceq p'$ and $p' \preceq p$ such that $p \neq p'$).

Such variants naturally appear when we study \emph{intersection of objectives} instead of a single objective as in Section~\ref{subsec:obj}:
\begin{itemize}
\item \emph{Intersection of homogeneous objectives}. For instance player~$1$ has $l$ reachability objectives $U_1, \ldots, U_l$ (instead of just one), and he wants to visit all the sets $U_1, \ldots, U_l$. 
\item \emph{Intersection of heterogeneous objectives}. In this more general case, player~$1$ has several objectives not necessarily of the same type. Let us imagine a situation where he has two quantitative objectives depending on two weight functions on the graph, like ensuring a threshold for the liminf of weights with respect to the first weight function and another threshold for the mean-payoff with respect to the second weight function.
\end{itemize}

In this context, for player~$1$, we consider a tuple $\bar f$ of payoff functions and a tuple $\bar w$ of weight functions (instead of a single payoff function $f$ defined from a single weight function $w$) such that each function $f_{k} : \Plays \to \mathbb R$ is defined from $w_{k}:E \to \mathbb Q$.\footnote{This tuple of payoff functions is used by player~$1$ contrarily to Definition~\ref{def:payoff} where function $f_i$ is used by player~$i$ for all $i \in \Pi$.} Tuples of payoffs $\bar p = \bar f(\rho)$ and $\bar p' = \bar f(\rho')$ are then compared using the usual ordering on tuples of reals: $\bar p \prec_{\sf ord} \bar p'$ iff $p_k \leq p'_k$ for all components $k$ and there exists $k$ such that $p_k < p'_k$ (the preference relation $\prec_{\sf ord}$ is not total). Let us mention some results 
first for quantitative objectives and then for Boolean objectives.

\subsubsection{Combination of quantitative objectives}

The threshold problem takes the following form: given a tuple $\bar \mu$ of rational thresholds, decide whether player~$1$ has a strategy $\sigma_1$ that ensures a payoff $\bar f(\rho)$ such that $\bar \mu \prec_{\sf ord} \bar \fctPayoff(\rho)$ for all plays $\rho$ consistent with~$\sigma_1$. 

\begin{theorem} \label{thm:meanpayoff} \cite{VelnerC0HRR15}
Let $(G,v_0)$ be an initialized two-player zero-sum game with homogeneous intersections of mean-payoff objectives. Then the threshold problem (in Problem~\ref{prob:twoplayers}) is~
\begin{itemize}
\item in {\sf NP $\cap$ co-NP} for functions $\SupMP$,
\item is {\sf co-NP}-complete for functions $\InfMP$.
\end{itemize}
In both cases, infinite memory is required for winning strategies of player~$1$ whereas uniform winning strategies are sufficient for player~$2$.
\end{theorem}

%

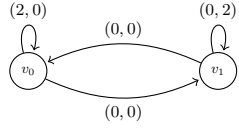
\begin{figure}[ht!] 
\begin{center}
\begin{tikzpicture}[initial text=,auto, node distance=2cm, shorten >=1pt]

\node[state, scale=0.6]              (1)                     {$v_0$};
\node[state, scale=0.6]   (2)    [right=of 1]     {$v_1$};

\path[->] (1) edge [bend right=25]          node[below, scale=0.7, black]    {$(0, 0)$}   (2)
edge  [loop above]            node[midway, scale=0.7, black]   {$(2,0)$}   ()

         (2) edge  [bend left=-25]               node[above, scale=0.7]           {$(0, 0)$}   (1)
         edge  [loop above]            node[midway, scale=0.7, black]   {$(0,2)$}   ();

\end{tikzpicture}

\caption{A one-player game with a pair of weight functions $\bar w$}
\label{fig:MP}
\end{center}
\end{figure}

This theorem indicates different behaviors for the functions $\SupMP$ and $\InfMP$. This is illustrated with the example of the initialized one-player game $(G,v_0)$ depicted in Figure~\ref{fig:MP}, where player~$1$ wants to ensure the intersection of two homogeneous objectives. It is shown in~\cite{VelnerC0HRR15} that for a pair of functions $\InfMP$, player~$1$ can ensure a threshold $(1,1)$, and that for a pair of functions $\SupMP$, he can ensure a threshold $(2,2)$ (which is impossible with $\InfMP$). In both cases infinite memory is necessary. Indeed recall that with a finite-memory strategy the produced play is a lasso $\rho = hg^\omega$ such that $\SupMP(\rho) = \InfMP(\rho)$ is the average weight of the cycle $g$. Here this average weight has the form $a \cdot (2,0) + b \cdot (0,0) + c \cdot (0,2) = (2a,2c)$, with $a+b+c = 1$ and $b > 0$. Clearly $(1,1) \not\prec_{\sf ord} (2a,2c)$ showing that player~$1$ is losing for threshold $(1,1)$ with finite-memory strategies. 

In~\cite{Velner15}, the author studies objectives equal to \emph{Boolean combinations} of inequalities $f_{k}(\rho) \sim \mu_k$, with $\sim$ $\in\{\leq, \geq \}$ and $f_{k} \in \{\SupMP,\InfMP\}$: deciding whether player~$1$ has a winning strategy in $(G,v_0)$ becomes undecidable. However, this problem remains decidable and is {\sf EXPTIME}-complete for CNF/DNF Boolean combinations of functions taken among $\{\Sup, \Inf, \LimSup, \LimInf, \WMP\}$~\cite{BruyereHR16}, where $\WMP$ is an interesting window variant of mean-payoff introduced in~\cite{Chatterjee0RR15}. The threshold problem is {\sf P}-complete (resp. {\sf EXPTIME}-complete) for a single $\WMP$ objective (resp. an intersection of $\WMP$ objectives)~\cite{Chatterjee0RR15}. Recall that it is in {\sf NP $\cap$ co-NP} for a single function $\SupMP$ or $\InfMP$ (see Theorem~\ref{thm:twoquantitative}).

\subsubsection{Combination of Boolean objectives}

Concerning one-player games, Boolean combinations of B\"uchi and co-B\"uchi objectives are introduced in~\cite{EmersonL87} as a generalization of Rabin and Streett objectives. It is proved that the non emptiness problem for this class of automata is {\sf NP}-complete (for a comparison see Theorem~\ref{thm:onePlayerOmegaReg}). 
Concerning two-player games, the intersection of homogeneous objectives is simple for safety, co-B\"uchi, Streett, and Muller cases. Indeed the intersection of safety (resp. co-B\"uchi, Streett, Muller) objectives is again a safety (resp. co-B\"uchi, Streett, Muller) objective. In the other cases, we have the following results to be compared with those of Theorem~\ref{thm:TwoPlayerOmegaReg}.

\begin{theorem} \label{thm:homogeneous}
Let $(G,v_0)$ be an initialized two-player zero-sum game with an intersection of homogeneous objectives. Then Problem~\ref{prob:twoplayers} is
\begin{itemize}
\item {\sf PSPACE}-complete for reachability objectives with finite-memory winning strategies for both players~\cite{FijalkowH13},
\item {\sf P}-complete for B\"uchi objectives with finite-memory (resp. uniform) winning strategies for player~$1$ (resp. player~$2$)~\cite{ChatterjeeDHL16},
\item {\sf co-NP}-complete for parity objectives with finite-memory (resp. uniform) winning strategies for player~$1$ (resp. player~$2$)~\cite{ChatterjeeHP07},
\item {\sf PSPACE}-complete for Rabin objectives with finite-memory winning strategies for both players.\footnote{We found no reference for this result. The {\sf PSPACE} membership (resp. the finite memory of the strategies) follows from~\cite{AlurTM03} (resp.~\cite{BruyereHR16}). In~\cite{AlurTM03}, games with a union of a Streett objective and a Rabin objective are shown to be {\sf PSPACE}-hard. It is thus also the case for games with a union of Streett objectives. By Martin's theorem, it follows that games with an intersection of Rabin objectives are {\sf PSPACE}-hard.}
\end{itemize}
\end{theorem}


Problem~\ref{prob:twoplayers} is {\sf PSPACE}-complete for heterogeneous intersections of reachability and B\"uchi objectives~\cite{BruyereHR16} as well as for Boolean combinations of B\"uchi objectives~\cite{AlurTM03,HunterD05}.


\subsubsection{Lexicographic and secure preferences}

For a tuple $\bar f$ of payoff functions defined from a tuple $\bar w$ of weight functions for player~$1$, let us mention two other natural preference relations $\prec$.
\begin{definition} \label{def:lexico}
Let $\bar p, \bar p'$ be two tuples of real payoffs.
\begin{itemize}
\item \emph{lexicographic preference}: $\bar p \prec_{\sf lex} \bar p'$ iff there exists $k$ such that $p_k < p'_k$ and $p_j = p'_j$ for all $j \leq k$. That is, player~$1$ prefers to first maximize the first component, then the second, then the third, etc (see for instance~\cite{BloemCHJ09}).
\item \emph{secure preference}: $\bar p \prec_{\sf sec} \bar p'$ iff either $p_1 < p'_1$ or \{$p_1 = p'_1$, $p_k \geq p'_k$ for all components $k > 1$, and there exists $k > 1$ such that $p_k > p'_k$\}. That is, player~$1$ prefers to first maximize the first component, and then to minimize all the other components (see for instance~\cite{Depril14}).
\end{itemize}
\end{definition}

The lexicographic preference is total whereas the secure preference is total only for \emph{pairs} (instead of tuples) of payoffs. In the latter case, we get a preference which is close to the lexicographic ordering: player~$1$ prefers to maximize the first component, and then to minimize the second one.
The secure preference is used in the notion of secure equilibrium discussed later in Section~\ref{subsec:secure}.

\begin{theorem} \label{thm:2PlayersOtherPref}
Let $(G,v_0)$ be an initialized two-player zero-sum game.
\begin{itemize}
\item Suppose that $\prec$ is the lexicographic preference $\prec_{\sf lex}$. Then the threshold problem (in Problem~\ref{prob:twoplayers}) for function $\InfMP$ is in {\sf NP $\cap$ co-NP} with uniform winning strategies for both players~\cite{BloemCHJ09}.
\item Suppose that $\prec$ is the secure preference $\prec_{\sf sec}$ on pairs of payoffs. Then the threshold problem (in Problem~\ref{prob:twoplayers}) is in {\sf NP $\cap$ co-NP} (resp. {\sf P}-complete) for functions $\SupMP$, $\InfMP$, and $\Disc^\lambda$ (resp. for  functions $\Sup$, $\Inf$, $\LimSup$, and $\LimInf$). Moreover both players have uniform (resp. positional) winning strategies for  functions $\LimSup$, $\LimInf$, $\SupMP$, $\InfMP$, and $\Disc^\lambda$ (resp. for functions $\Sup$ and $\Inf$)~\cite{BMR14}.
\end{itemize}
\end{theorem}

In this theorem, it is supposed that the components $f_{k}$ of $\bar f$ are all of the same type (for instance they are all limsup functions); and some results stating the existence of uniform winning strategies can be established thanks to Theorem~\ref{thm:mixing}. Notice that the secure preference is limited to pairs of payoffs in a way to be total, which is a necessary condition when dealing with values. Notice also that the authors in~\cite{BloemCHJ09} consider liminf average of the weight \emph{vector} under lexicographic ordering whereas the authors in~\cite{BMR14} consider the secure ordering of components where each \emph{component} is the liminf average value.

The threshold problem is studied in~\cite{BouyerBMU15} in a general context: the players can use various preorders  (like the lexicographic preference, a preorder given by a Boolean circuit, etc), the players play concurrently and not in a turned-based way, and the objectives are Boolean as in Definition~\ref{def:omega-reg}.

\section{Multi-player non zero-sum games} \label{sec:multi}

In multi-player non zero-sum games, the different players $i \in \Pi$ are not necessarily antagonistic, they have their own payoff functions $f_i$ and preference relations $\prec_i$. Each of them follows a strategy $\sigma_i$, the resulting strategy profile $(\sigma_i)_{i \in \Pi}$ induces a play that should be satisfactory to all players. As explained in Section~\ref{subsec:probl} (see Definition~\ref{def:NE}), a classical solution profile is the notion of NE, where no player has an incentive to deviate when the other players stick to their own strategies. It is proved in~\cite{ChatterjeeMJ04,GU08} that there exists an NE in every initialized multi-player non zero-sum game with Borel Boolean objectives. We go further by presenting in this section additional existence results for quantitative games and some known results for NEs as a solution to Problem~\ref{prob:multiplayers} (threshold problem and constraint problem). As in Section~\ref{sec:two}, we focus on general approaches.

\subsection{Characterization of outcomes of NE} 

Given a multi-player non zero-sum game $G$ and an initial vertex $v_0$, we begin by a \emph{characterization} of plays $\rho \in \Plays(v_0)$ that are the outcome of an NE $(\sigma_i)_{i \in \Pi}$ in $(G,v_0)$. It will imply the existence of NE in large classes of games (see Corollaries~\ref{cor:existNE} and~\ref{cor:existNEbis}), and will be useful for the study of Problem~\ref{prob:multiplayers} (see Theorems~\ref{thm:NEprob2} and~\ref{thm:NEprob2quant}). This characterization is related to a \emph{family of two-player zero-sum games} $G_i$, one for each $i \in \Pi$, associated with $G$ and defined as follows. \emph{(i)} The game $G_i$ has the same arena as $G$, \emph{(ii)} the two players are player~$i$ (player~$1$) and player~$-i$ (player~$2$) formed by the \emph{coalition} of the other players $j \in \Pi \setminus \{i\}$, \emph{(iii)} the payoff function of player~$i$ is equal to $f_i$ and his preference relation is equal to $\prec_i$\footnote{Recall that the payoff function and the preference relation of the second player do not matter in two-player zero-sum games.}. For all $v \in V$, when it exists, we denote by $\val_i(v)$ the value of vertex $v$ in game $G_i$, and by $\tau_i^v$, $\tau_{-i}^v$ the related optimal strategies for players $i, -i$ respectively in $(G_i,v)$ (see Definition~\ref{def:value}).

\begin{proposition} \label{prop:caracNE}
Let $G$ be a multi-player non zero-sum game such that for all $i \in \Pi$,
\begin{itemize}
\item the payoff function $\fctPayoff_i$ is prefix-linear, and
\item in the game $G_i$, all vertices has a value.
\end{itemize}
Then $\rho = \rho_0\rho_1 \ldots \in \Plays(v_0)$ is the outcome of an NE in $(G,v_0)$ iff $\val_i(\rho_k) \preceq_i \fctPayoff_i(\rho_{\geq k})$ for all $i \in \Pi$ and all $k \in \mathbb N$ such that $\rho_k \in V_i$.
\end{proposition}

The condition of this proposition asks that for all $k$, if vertex $\rho_k$ is controlled by player~$i$, then in the two-player zero-sum game $G_i$, its value is less preferred or equal to the payoff of the suffix $\rho_{\geq k}$.
The proposed characterization appears under various particular forms, for instance in \cite{BMR14,GU08,UmmelsW11}. It is here given under two general conditions already studied in Section~\ref{subsec:two}. Recall that almost all the payoff functions considered in Section~\ref{subsec:obj} are prefix-linear and that for all the related two-player zero-sum games $G_i$, the vertices have a value. Notice that when $f_i$ is prefix-independent, condition $\val_i(\rho_k) \preceq_i \fctPayoff_i(\rho_{\geq k})$ for all $k \in \mathbb N$ with $\rho_k \in V_i$ simplifies in $\max \{\val_i(\rho_k) \mid k \in \mathbb N, \rho_k \in V_i \} \preceq_i \fctPayoff_i(\rho)$ (the maximum exists since $V_i$ is finite). 

\addtocounter{example}{-26} 
\begin{example}[continued]
An example of NE with outcome $\rho = v_0v_1v_3^\omega$ was given in Example~\ref{ex:first-ex} for the initialized game $(G,v_0)$ of Figure~\ref{fig:gameAbstract}. Let us verify that $\rho$ satisfies the characterization of Proposition~\ref{prop:caracNE}. Recall that both players use the same payoff function $f$ that is prefix-independent. The values of $G_1$ were computed in Example~\ref{ex:values}: $\val_1(v_0) = \val_1(v_1) = \val_1(v_2) = p_2$ and $\val_1(v_3) = p_3$. Similarly one can compute the values of $G_2$: $\val_2(v_0) = \val_2(v_2) = p_2$ and $\val_2(v_1) = \val_2(v_3) = p_3$. One checks that $\max \{\val_i(\rho_k) \mid k \in \mathbb N, \rho_k \in V_i \} \preceq_i \fctPayoff(\rho) = p_3$, for $i = 1,2$.
\end{example}
\addtocounter{example}{+25} 

The proof of Proposition~\ref{prop:caracNE} is easy to establish. 

Firstly suppose that $\rho$ is the outcome of an NE $(\sigma_i)_{i \in \Pi}$ and that there exist $i \in \Pi$ and $k \in \mathbb N$ with $\rho_k \in V_i$ such that $\fctPayoff_i(\rho_{\geq k}) \prec_i \val_i(\rho_k)$. Let us show that player~$i$ has a profitable deviation $\sigma'_i$ with respect to $(\sigma_i)_{i \in \Pi}$ in contradiction with $(\sigma_i)_{i \in \Pi}$ being an NE. The strategy $\sigma'_i$ consists in playing according to $\sigma_i$ until producing $\rho_{\leq k}$ and from $\rho_k$ in playing according to his optimal strategy $\tau_i^{\rho_k}$ (in $(G_i,\rho_k)$). The payoff of the resulting play $\pi$ from $\rho_k$ is such that $\val_i(\rho_k) \preceq_i \fctPayoff_i(\pi)$ by optimality of $\tau_i^{\rho_k}$, and thus $f_i(\rho_{\geq k}) \prec_i \fctPayoff_i(\pi)$. From prefix-linearity of $\fctPayoff_i$ it follows that $\fctPayoff_i(\rho) = \fctPayoff_i(\rho_{< k}\rho_{\geq k}) \prec_i \fctPayoff_i(\rho_{< k}\pi)$ as required.

Secondly suppose that $\val_i(\rho_k) \preceq_i \fctPayoff_i(\rho_{\geq k})$ for all $i \in \Pi$ and all $k \in \mathbb N$ such that $\rho_k \in V_i$. We are going to construct an NE by using a well-known method in classical game theory that is used in the proof of the Folk Theorem in repeated games~\cite{Osborne94}. We define a strategy profile $(\sigma_i)_{i \in \Pi}$ that produces $\rho$ as outcome, and as soon as some player~$i$ deviates from $\rho$, say at vertex $\rho_k$, all the other players (as a coalition) punish him by playing from $\rho_k$ the optimal strategy $\tau_{-i}^{\rho_k}$ (in $(G_i,\rho_k)$). Let us show that $(\sigma_i)_{i \in \Pi}$ is an NE. Let $\sigma'_i$ be a deviating strategy from $\sigma_i$ for player~$i$, and let $\rho'$ be the outcome of the strategy profile $(\sigma'_i,\sigma_{-i})$. Consider the longest common prefix $\rho_{\leq k}$ of $\rho$ and $\rho'$. Then $\rho_k \in V_i$ and by optimality of $\tau_{-i}^{\rho_k}$, we get $\fctPayoff_i(\rho'_{\geq k}) \preceq_i \val_i(\rho_k)$ and thus $\fctPayoff_i(\rho'_{\geq k}) \preceq_i \fctPayoff_i(\rho_{\geq k})$. From prefix-linearity of $\fctPayoff_i$ it follows that $\fctPayoff_i(\rho') \preceq_i \fctPayoff_i(\rho)$ showing that $\sigma'_i$ is not a profitable deviation for player~$i$. 

Notice that in this proof, the first (resp. second) implication only requires condition (\ref{eq:2}) (resp. (\ref{eq:1})) of prefix-linearity of $\fctPayoff_i$. The next corollary follows from this observation and Proposition~\ref{prop:caracNE}.

\begin{corollary} \label{cor:existNE} \cite{DePril13}
Let $G$ be a multi-player non zero-sum game such that for all $i \in \Pi$,
\begin{itemize}
\item the payoff function $\fctPayoff_i$ satisfies $\fctPayoff_i(\rho) \preceq_i \fctPayoff_i(\rho') \Rightarrow \fctPayoff_i(h \rho) \preceq_i \fctPayoff_i(h \rho')$ for all $h\rho, h\rho' \in \Plays$, and
\item each game $G_i$ has uniform optimal strategies for both players.
\end{itemize}
Then there exists a finite-memory NE in each initialized game $(G,v_0)$.
\end{corollary}
 
This corollary is a generalization of a theorem\footnote{In \cite{TJS13}, one hypothesis is missing: the required optimal strategies must be uniform.} given in \cite{TJS13,DePril13} for the existence of NEs in games equipped with payoff functions $f_i : \Plays \to \mathbb R$, $i \in \Pi$. The proof of Corollary~\ref{cor:existNE} is as follows. Let us consider the play $\rho \in \Plays(v_0)$ produced by the players when each player~$i$ plays according to his optimal strategy $\tau_i$ in $(G_i,v_0)$ ($\tau_i^v =  \tau_i$ for all vertices $v$ since it is uniform). By construction, $\rho$ is the outcome of an NE because it satisfies the characterization of Proposition~\ref{prop:caracNE}. Notice that $\rho$ is a simple lasso since each $\tau_i$, $i \in \Pi$, is uniform. Therefore the strategies of the constructed NE are finite-memory with a small memory size bounded by $|V| + |\Pi|$ to remember this lasso and the first player who deviates from $\rho$.

The existence of an NE is also guaranteed in the following corollary that does not require optimal strategies that are uniform, but in counterpart requires payoff functions that are prefix-independent. 

\begin{corollary} \label{cor:existNEbis} \cite{DePril13}
Let $G$ be a multi-player non zero-sum game such that for all $i \in \Pi$,
\begin{itemize}
\item the payoff function $\fctPayoff_i$ is prefix-independent, and
\item each game $G_i$ has (resp. finite-memory) optimal strategies for both players.
\end{itemize}
Then there exists an (resp. finite-memory) NE in each initialized game $(G,v_0)$.
\end{corollary}

This is a generalization of a result given in~\cite{DePril13} for games equipped with payoff functions $f_i : \Plays \to \mathbb R$, $i \in \Pi$. The proof is as follows: under the hypotheses of Corollary~\ref{cor:existNEbis}, one can show that there exist optimal strategies ${\tau}_i^{v_0}$ in $(G_i,v_0)$, $i \in \Pi$, such that for all plays $\rho \in \Plays(v_0)$ consistent with ${\tau}_i^{v_0}$, we have $\max \{\val_i(\rho_k) \mid k \in \mathbb N, \rho_k \in V_i \} \preceq_i \fctPayoff_i(\rho)$. Then as in Corollary~\ref{cor:existNE}, we consider the play $\rho \in \Plays(v_0)$ obtained when each player~$i$ plays according to his optimal strategy ${\tau}_i^{v_0}$. As each $\fctPayoff_i$ is prefix-independent, $\rho$ satisfies the characterization of Proposition~\ref{prop:caracNE}.

From Corollaries~\ref{cor:existNE} and~\ref{cor:existNEbis}, it follows that there exists an NE (which can be constructed) in every game of Section~\ref{subsec:obj}; the case of Boolean (resp. quantitative) game is proved in~\cite{ChatterjeeMJ04,GU08} (resp. in~\cite{TJS13,DePril13}). The existence of an NE in discounted sum games can be obtained in a second way: the function~$\Disc^\lambda$ is continuous and all games with real-valued continuous payoff functions always have an NE~\cite{Fudenberg83,Harris85}. Notice that the two previous corollaries allow mixing the types of functions $f_i$, like for instance $f_1$ associated with a B\"uchi objective, a limsup function $f_2$, a mean-payoff function $f_3$, etc.  

Conditions generalizing those of Corollaries~\ref{cor:existNE} and~\ref{cor:existNEbis} are given in~\cite{RouxP16} that guarantee the existence of a finite-memory NE. Moreover, for most of the given conditions counterexamples are provided that show that they cannot be dispensed with. 

\subsection{Solution to Problem~\ref{prob:multiplayers}}

In this section we study how to solve Problem~\ref{prob:multiplayers} for NEs (threshold problem and constraint problem). The characterization given in Proposition~\ref{prop:caracNE} provides a general approach to solve this problem. Indeed consider the case of initialized games $(G,v_0)$ with prefix-independent payoff functions $f_i$ and such that the vertices of each game $G_i$, $i \in \Pi$, has a value. Then given two tuples of bounds $(\mu_i)_{i \in \Pi}, (\nu_i)_{i \in \Pi}$, we simply have to check whether there exists a play $\rho \in \Plays(v_0)$ such that for all $i \in \Pi$,  
\begin{eqnarray} \label{eq:constraint}
\max \{\val_i(\rho_k) \mid \rho_k \in V_i \} \preceq_i \fctPayoff_i(\rho) \mbox{ and } 
\mu_i \preceq_i \fctPayoff_i(\rho) \mbox{ (resp. } \mu_i \preceq_i \fctPayoff_i(\rho) \preceq_i \nu_i). 
\end{eqnarray}

Thanks to this general approach or variations based on Proposition~\ref{prop:caracNE}, Problem~\ref{prob:multiplayers} is solved for B\"uchi, co-B\"uchi, Streett, and parity games in~\cite{Ummels08}, and for the other Boolean games in~\cite{ConduracheFGR16}. 

\begin{theorem} \label{thm:NEprob2} \cite{ConduracheFGR16,Ummels08}
Let $(G,v_0)$ be an initialized multi-player non zero-sum Boolean game. Then Problem~\ref{prob:multiplayers} is
\begin{itemize}
\item is {\sf P}-complete for B\"uchi and Muller\footnote{We found no reference for Muller objectives. A sketch of proof is given in the appendix. Problem~\ref{prob:multiplayers} is {\sf PSPACE}-complete for the colored variant of Muller objectives~\cite{ConduracheFGR16}.} games,
\item {\sf NP}-complete for reachability, safety, co-B\"uchi, parity and Streett games,
\item in ${\sf P^{NP}}$, and {\sf NP}-hard, {\sf co-NP}-hard for Rabin games.
\end{itemize}
\end{theorem}

Let us explain the proof of {\sf NP} membership for parity games and the constraint problem with bounds $(\mu_i)_{i \in \Pi}, (\nu_i)_{i \in \Pi}$. As each $G_i$ is a two-player zero-sum parity game, recall that the constraint problem is in {\sf NP $\cap$ co-NP} with uniform winning strategies for both players (see Theorem~\ref{thm:TwoPlayerOmegaReg}). The required algorithm in {\sf NP} is as follows. \emph{(i)} For all $i \in \Pi$,  in the game $G_i$, guess a subset $U_i \subseteq V$ of vertices and two uniform strategies $\tau_i, \tau_{-i}$ for players $i, -i$ respectively (intuitively we guess $U_i = \{v \in V \mid \val_i(v)=1\}$ and $V \setminus U_i = \{v \in V \mid \val_i(v)=0\}$). Check in polynomial time\footnote{Recall our comment after Theorem~\ref{thm:twoquantitative}.} that $\tau_i$ is a winning strategy for player~$i$ for the constraint problem in each $(G_i,v)$ with $v \in U_i$ and that $\tau_{-i}$ is a winning strategy for player~$-i$ for the opposite objective in each $(G_i,v)$ with $v \in V \setminus U_i$.
\emph{(ii)} Then for all $i \in \Pi$, we guess $r_i \in V_i$ (intuitively we guess $r_i$ such that $\val_i(r_i) = \max \{\val_i(\rho_k) \mid k \in \mathbb N, \rho_k \in V_i \}$ for the required play $\rho$). 
Construct in polynomial time a one-player game $G'$ from $G$ such that each set $V_i$ of vertices is limited to $\{v \in V_i \mid \val_i(v) \leq \val_i(r_i) \}$ and the unique player is formed by the coalition of all players $i \in \Pi$.
\emph{(iii)} By (\ref{eq:constraint}) it remains to check whether there exists a play $\rho$ in $(G',v_0)$ such that for all $i \in \Pi$, $\val_i(r_i) \leq \fctPayoff_i(\rho)$ and $\mu_i \leq \fctPayoff_i(\rho) \leq \nu_i$. Recall that the existence of plays satisfying certain constraints in one-player games was studied in Section~\ref{subsec:one}, see Theorem~\ref{thm:onePlayerOmegaReg}. Here we are faced with the existence of a play in game with an intersection of parity objectives which can be checked in polynomial time by~\cite{EmersonL87}.

Problem~\ref{prob:multiplayers} can be similarly solved for quantitative games. 

\begin{theorem} \label{thm:NEprob2quant} \cite{ConduracheFGR16,Ummels08,UmmelsW11}
Let $(G,v_0)$ be an initialized multi-player non zero-sum quantitative (except discounted sum) game. Then Problem~\ref{prob:multiplayers} is
\begin{itemize}
\item {\sf P}-complete for limsup games,
\item {\sf NP}-complete for supremum, infimum, liminf, mean-payoff $\InfMP_i$, and mean-payoff $\SupMP_i$ games.
\end{itemize}
\end{theorem}

The case of supremum, infimum, limsup and liminf games is equivalent to the case of reachability, safety, B\"uchi and co-B\"uchi games presented in Theorem~\ref{thm:NEprob2}, whereas the case of mean-payoff games is studied in~\cite{UmmelsW11}. The proof of {\sf NP} membership for mean-payoff games is based on the approach (\ref{eq:constraint}), and is similar to the one given above for parity games.
%
%
%
The case of discounted sum games is open. Indeed it is proved in~\cite{BMR14} that Problem~\ref{prob:Boker} reduces to Problem~\ref{prob:multiplayers} with the discounted sum function.\footnote{The reduction is given for another kind of solution profile but it also works for NEs.} 

Problem~\ref{prob:multiplayers} is studied in~\cite{BouyerBMU15} in a general context: the players can use various preorders, they play concurrently and not in a turned-based way, and the objectives are Boolean as in Definition~\ref{def:omega-reg}. The general approach proposed in~\cite{BouyerBMU15} is different from the one of Proposition~\ref{prop:caracNE}.
 
\subsection{Other solution profiles}

In this section, we present some other solution profiles. Indeed the notion of NE has several drawbacks: \emph{(i)} Each player is selfish since he is only concerned with his own payoff, and not with the payoff of the other players. \emph{(ii)} An NE does not take into account the sequential nature of games played on graphs. We illustrate these drawbacks in the following two examples of quantitative game.

\begin{example} Consider the two-player quantitative game of Figure~\ref{fig:notsecure} such that $f_i = \LimSup_i$ for $i = 1,2$. The strategy profile depicted with thick edges is an NE. Notice that player~$1$ could decide to deviate at $v_0$ by moving to $v_2$. Indeed he then keeps the same payoff of~$1$ but also decreases the payoff of player $2$ (from~$2$ to $1$) which is bad for player~$2$. To avoid such a drawback, we will introduce hereafter the concept of secure equilibrium, where each player take cares of his own payoff as well as the payoff of the other players (but in a negative way).

Consider now the game of Figure~\ref{fig:notSPE} where the weights of the loops have been modified. The depicted strategy profile is again an NE. Player~$1$ has no incentive to deviate at $v_0$ due to the threat of player~$2$: player~$1$ will receive a payoff of $0 < 1$. Such a threat of player~$2$ is non credible because in the subgame induced by $v_2, v_3, v_4$, at vertex $v_2$, it is more rational for player~$2$ to move to $v_4$ to get a payoff of $2$ instead of going to $v_3$ where he only receives a payoff of $1$. To avoid such a drawback, we will introduce hereafter the concept of subgame perfect equilibrium that takes into account rational behaviors of the players in all subgames of the initial game.

\begin{figure}[h!]
\begin{minipage}[c]{.46\linewidth}
\begin{center}
\begin{tikzpicture}[initial text=,auto, node distance=1cm, shorten >=1pt] 

\node[state, scale=0.6]               (0)                          {$v_0$};
\node[state, scale=0.6]               (1)    [below left=of 0]     {$v_1$};
\node[state, rectangle, scale=0.6]    (2)    [below right=of 0]    {$v_2$};
\node[state, scale=0.6]               (3)    [below left=of 2]     {$v_3$};
\node[state, scale=0.6]               (4)    [below right=of 2]    {$v_4$};

\node (fictif) [above=4 mm of 0]  {};

\path[->] (0)  edge [thick, black]                     node[left, scale=0.7]           {$(0,0)$}          (1)
               edge                      node[right, scale=0.7]          {$(0,0)$}         (2)

          (1)  edge  [loop below, thick, black]        node[midway, scale=0.7]         {$(1,2)$}            ()

          (2)  edge [thick, black]                     node[left, scale=0.7]           {$(0,0)$}          (3)
               edge                      node[right, scale=0.7]          {$(0,0)$}         (4)

          (3)  edge  [loop below, thick, black]        node[midway, scale=0.7, black]  {$(1,1)$}            ()

          (4)  edge  [loop below, thick, black]        node[midway, scale=0.7, black]  {$(2,0)$}            ();

\end{tikzpicture}
\caption{An NE that is not a secure equilibrium \label{fig:notsecure}}
\end{center}
\end{minipage}
\hfill
\begin{minipage}[c]{.46\linewidth}
\begin{center}
\begin{tikzpicture}[initial text=,auto, node distance=1cm, shorten >=1pt] 

\node[state, scale=0.6]               (0)                          {$v_0$};
\node[state, scale=0.6]               (1)    [below left=of 0]     {$v_1$};
\node[state, rectangle, scale=0.6]    (2)    [below right=of 0]    {$v_2$};
\node[state, scale=0.6]               (3)    [below left=of 2]     {$v_3$};
\node[state, scale=0.6]               (4)    [below right=of 2]    {$v_4$};

\node (fictif) [above=4 mm of 0]  {};

\path[->] (0)  edge [thick, black]                     node[left, scale=0.7]           {$(0,0)$}          (1)
               edge                      node[right, scale=0.7]          {$(0,0)$}         (2)

          (1)  edge  [loop below, thick, black]        node[midway, scale=0.7]         {$(1,1)$}            ()

          (2)  edge [thick, black]                      node[left, scale=0.7]           {$(0,0)$}          (3)
               edge                     node[right, scale=0.7]          {$(0,0)$}         (4)

          (3)  edge  [loop below, thick, black]        node[midway, scale=0.7, black]  {$(0,1)$}            ()

          (4)  edge  [loop below, thick, black]        node[midway, scale=0.7, black]  {$(3,2)$}            ();

\end{tikzpicture}
\caption{An NE that is not a subgame perfect equilibrium \label{fig:notSPE}}
\end{center}
\end{minipage}

\end{figure}
\end{example}

\subsubsection{Secure equilibria} \label{subsec:secure}

The notion of \emph{secure equilibrium} (SE) is introduced in~\cite{CHJ06} for two-player non zero-sum games. The idea of an SE is that no player has an incentive to deviate in the following sense: he will not be able to increase his payoff, and keeping the same payoff he will not be able to decrease the payoff of the other player. An SE can thus be seen as a contract between the two players which strengthens cooperation: if a player chooses another strategy that is not harmful to himself, then this cannot harm the other player if the latter follows the contract.

The definition of an SE is given in the context of games equipped with payoff functions $f_i : \Plays \to \mathbb R$, $i \in \Pi$. It uses the notion of secure preference introduced in Section~\ref{subsec:preorders} (see Definition~\ref{def:lexico}\footnote{The definition was given for player~$1$.}). Let us recall the \emph{secure preference} $\prec_{{\sf sec},i}$ for player~$i$: given $\bar p = (f_i(\rho))_{i \in \Pi}, \bar p' = (f_i(\rho'))_{i \in \Pi}$, we have $\bar p \prec_{{\sf sec},i} \bar p'$ iff either $p_i < p'_i$ or \{$p_i = p'_i$, $p_k \geq p'_k$ for all components $k \neq i$, and there exists $k \neq i$ such that $p_k > p'_k$\}. Hence player~$i$ prefers to increase his own payoff, and in case of equality to decrease the payoffs of all the other players. This preference relation is not total except when there are only two players. 

The definition of an SE is very close to the one of NE (see Definition~\ref{def:NE}). The only difference is that it uses the secure preference:

\begin{definition} \label{def:SE}
Given an initialized game $(G, v_0)$, a strategy profile $(\sigma_i)_{i \in \Pi}$ is a \emph{secure equilibrium} if  
$$(\fctPayoff_i(\out{(\sigma_i)_{i \in \Pi}}_{v_0}))_{i \in \Pi} \nprec_{{\sf sec},i} (\fctPayoff_i(\out{\sigma'_i, \sigma_{-i}}_{v_0}))_{i \in \Pi}$$
for all players $i \in \Pi$ and all strategies $\sigma'_i$ of player~$i$.
\end{definition}

\addtocounter{example}{-2} 
\begin{example}[continued]
The strategy profile of Figure~\ref{fig:notsecure} is not an SE because player~$1$ has a profitable deviation if at $v_0$ he chooses to move to $v_2$: $(1,2) \prec_{{\sf sec},1}(1,1)$.
\end{example}
\addtocounter{example}{+1} 

By definition, every SE is an NE but the converse is false as shown in the previous example. It is proved in~\cite{CHJ06} that every two-player non zero-sum game with Borel Boolean objectives has an SE; this result is generalized to multi-player games in~\cite{Depril14}.

Let us turn to quantitative games such that all the players have the same type of payoff function $f_i$. General hypotheses are provided in~\cite{Depril14} that guarantee the existence of an SE in quantitative games, except for functions $\SupMP_i$ and $\InfMP_i$. Thanks to Corollary~\ref{cor:existNE} and Theorem~\ref{thm:2PlayersOtherPref}, 
for two-player\footnote{A restriction to two-player games is necessary to deal with a secure preference that is total.} quantitative games (now including functions $\SupMP_i$ and $\InfMP_i$), there exists such an  SE that is finite-memory~\cite{BMR14}. Moreover, with the same general approach (\ref{eq:constraint}) described previously for NEs, Problem~\ref{prob:multiplayers} is solved as follows for SEs.

\begin{theorem} \label{thm:Secureprob2quant} \cite{BMR14}
Let $(G,v_0)$ be an initialized two-player non zero-sum quantitative (except discounted sum) game. Then Problem~\ref{prob:multiplayers} for SEs is
\begin{itemize}
\item {\sf P}-complete for supremum, infimum, limsup, and liminf functions,
\item in {\sf NP $\cap$ co-NP} for functions $\SupMP_i$ and $\InfMP_i$.
\end{itemize}
\end{theorem}


The case of discounted sum function is open since it is proved in~\cite{BMR14} that Problem~\ref{prob:Boker} reduces to Problem~\ref{prob:multiplayers} with $\Disc^\lambda$. The complexity class of the problem of deciding whether, in an initialized two-player parity game $(G,v_0)$, there exists an SE with payoff respectively equal to $(0,0)$, $(0,1)$, $(1,0)$, and $(1,1)$, is studied in~\cite{CHJ06,GU08}.

\subsubsection{Subgame perfect equilibria}

A solution profile that avoids incredible threats by taking into account the sequential nature of games played on graphs is the notion of \emph{subgame perfect equilibrium} (SPE)~\cite{selten}. For being an SPE, a strategy profile is not only required to be an NE from the initial vertex but after every possible history of the game. 

Before giving the definition of an SPE, we need to introduce the following concepts for an initialized  game $(G,v_0)$ with payoff functions $f_i$ and preference relations $\prec_i$, for all $i \in \Pi$. Given a history $hv \in \Hist(v_0)$, the \emph{subgame} $(\Sub{G}{h},v)$ of $(G,v_0)$ is the initialized game with payoff functions $\Sub{f_i}{h}$, $i \in \Pi$, such that $\Sub{\fctPayoff_i}{h}(\rho) = \fctPayoff_i(h\rho)$ for all plays $\rho \in \Plays(v)$ (the preference relation of player~$i$ is his preference $\prec_i$ in $G$).  Given a strategy $\sigma_i$ for player $i$ in $(G,v_0)$, the strategy $\Sub{\sigma_i}{h}$ in $(\Sub{G}{h}, v)$ is defined as $\Sub{\sigma_i}{h}(h') = \sigma_i(hh')$ for all $h' \in \Hist_i(v)$. 

\begin{definition} \label{def:SPE}
Given an initialized game $(G, v_0)$, a strategy profile $(\sigma_i)_{i \in \Pi}$ is a \emph{subgame perfect equilibrium} if $(\Sub{\sigma_i}{h})_{i \in \Pi}$ is an NE in each subgame $(\Sub{G}{h}, v)$ of $(G,v_0)$ with $hv \in \Hist(v_0)$. 
\end{definition}

\addtocounter{example}{-4} 
\begin{example}[continued]
The strategy profile $(\sigma_1,\sigma_2)$ of Figure~\ref{fig:notSPE} is not an SPE because in the subgame $(\Sub{G}{v_0},v_2)$, player~$2$ has a profitable deviation with respect to $(\Sub{\sigma_1}{v_0},\Sub{\sigma_2}{v_0})$ if at $v_2$ he chooses to move to $v_4$.
\end{example}
\addtocounter{example}{+3} 

By definition, every SPE is an NE but the converse is false as shown in the previous example.  A well-known result is the existence of an SPE in every initialized game $(G,v_0)$ such that its arena is a tree rooted at $v_0$\footnote{In this particular context, plays are finite paths.}~\cite{kuhn53}. The SPE is constructed backwards from the leaves to the initial vertex $v_0$ in the following way. Suppose that the current vertex $v$ is controlled by player~$i$, and that for each son $v'$ of $v$ one has already constructed an SPE $(\sigma_i^{v'})_{i \in \Pi}$ in the subtree rooted at $v'$. Then player~$i$ chooses the edge $(v,v')$ such that $(\sigma_i^{v'})_{i \in \Pi}$ has the best outcome with respect to his preference relation $\prec_i$. The resulting strategy profile $(\sigma^{v}_i)_{i \in \Pi}$ is an SPE in the subtree rooted at $v$.

It is proved in~\cite{Ummels06} that there exists an SPE in every multi-player non zero-sum game with Borel Boolean objectives, and that in case of $\omega$-regular objectives there exists one that is finite-memory. Existence of an SPE also holds for games with continuous real-valued payoff functions~\cite{Fudenberg83,Harris85} (this is also holds when the functions are upper-semicontinuous (resp. lower-semicontinuous) and with finite range~\cite{Flesch10} (resp. \cite{Purves11})).

For subgame perfect equilibria, we are not aware of a characterization like the one in Proposition~\ref{prop:caracNE}. Therefore a solution to Problem~\ref{prob:multiplayers} for SPEs needs a different approach. Few solutions are known: this problem is in {\sf EXPTIME} for Rabin games~\cite{Ummels06} and is {\sf NP}-hard for co-B\"uchi games~\cite{GU08}.

Whereas NEs exist for large classes of games, see Corollaries~\ref{cor:existNE} and~\ref{cor:existNEbis}, SPEs fail to exist even in simple games like the one of Figure~\ref{fig:gameAbstract}~\cite{SV03}. Variants of SPE, \emph{weak SPE} and \emph{very weak SPE}, have thus been proposed in~\cite{BBMR15} as interesting alternatives. In a weak SPE (resp. very weak SPE), a player who deviates from a strategy $\sigma$ is allowed to use deviating strategies that differ from $\sigma$ on a \emph{finite number} of histories only (resp. only on the \emph{initial vertex}). 
Deviating strategies that only differ on the initial vertex are a well-known notion that for instance appears in the proof of Kuhn's theorem~\cite{kuhn53} with the one-step deviation property. By definition, every SPE is a weak SPE, and every weak SPE is a very weak SPE. Weak SPE and very weak SPE are equivalent notions, but this is not true for SPE and weak SPE~\cite{BBMR15}. 

%

The following theorem gives two general conditions such that each of them separately guarantees the existence of a weak SPE.

\begin{theorem} \cite{Bruyere0PR17} \label{thm:roux}
Let $G$ be a multi-player non zero-sum game such that
\begin{itemize}
\item either each payoff function $\fctPayoff_i$, $i \in \Pi$, is prefix-independent,
\item or each $\fctPayoff_i$, $i \in \Pi$, has a finite range.
\end{itemize}
Then there exists a weak SPE in each initialized game $(G,v_0)$.
\end{theorem}

This theorem has to be compared with Corollary~\ref{cor:existNEbis} that gives general conditions for the existence of an NE, one of them being prefix-independence of $f_i$, $i \in \Pi$. This latter condition is here enough to guarantee the existence of a weak SPE (the existence of an SPE is not possible as mentioned before with the game of Figure~\ref{fig:gameAbstract}~\cite{SV03}). 
It follows from Theorem~\ref{thm:roux} that there exists a weak SPE in all the Boolean and quantitative games of Section~\ref{subsec:obj} (except for the case of discounted sum payoff that is neither prefix-independent nor with finite range). 

In addition to SEs and (weak) SPEs, other solution profiles have been recently proposed, like Doomsday equilibria in~\cite{ChatterjeeDFR17}, robust equilibria in~\cite{Brenguier16}, and equilibria using admissible strategies in~\cite{BrenguierRS17}. We also refer the reader to the survey~\cite{BrenguierCHPRRS16}.

\section{Conclusion} \label{sec:conclusion}

In this invited contribution, we gave an overview of classical as well as recent results about the threshold and constraint problems for games played on graphs. Solutions to these problems are winning strategies in case of two-player zero-sum games, and equilibria in case of multi-player non zero-sum games. We tried to present a unified approach through the notion of games equipped with a payoff function and a preference relation for each player, in a way to include classes of Boolean games and quantitative games that are usually studied. We also focussed on general approaches from which one can derived several different results: a criterium that guarantees the existence of uniform optimal strategies in two-player zero-sum games, and a characterization of plays that are the outcome of an Nash equilibrium in multi-player non zero-sum games. Several illustrative examples were provided as well as some intuition on the proofs when they are simple.

\section*{Acknowledgments}

We would like to thank Patricia Bouyer, Thomas Brihaye, Emmanuel Filiot, Hugo Gimbert, Quentin Hautem, Micka\"el Randour, and Jean-Fran\c cois Raskin for their useful discussions and comments that helped us to improve the presentation of this article.

\bibliographystyle{plain}
\bibliography{biblio}

\begin{thebibliography}{10}

\bibitem{AlurTM03}
Rajeev Alur, Salvatore {La Torre}, and P.~Madhusudan.
\newblock Playing games with boxes and diamonds.
\newblock In {\em {CONCUR} Proceedings}, volume 2761 of {\em Lecture Notes in
  Comput. Sci.}, pages 127--141. Springer, 2003.

\bibitem{And06}
Daniel Andersson.
\newblock An improved algorithm for discounted payoff games.
\newblock In {\em ESSLLI Student Session}, pages 91--98, 2006.

\bibitem{Beeri}
Catriel Beeri.
\newblock On the membership problem for functional and multivalued dependencies
  in relational databases.
\newblock {\em ACM Trans. Database Syst.}, 5(3), September 1980.

\bibitem{BR16}
Val\'erie Berth\'e and Michel Rigo, editors.
\newblock {\em Combinatorics, Words and Symbolic Dynamics}, volume 135.
\newblock Cambridge University Press, 2016.

\bibitem{BloemCHJ09}
Roderick Bloem, Krishnendu Chatterjee, Thomas~A. Henzinger, and Barbara
  Jobstmann.
\newblock Better quality in synthesis through quantitative objectives.
\newblock In {\em {CAV} Proceedings}, volume 5643 of {\em Lecture Notes in
  Comput. Sci.}, pages 140--156. Springer, 2009.

\bibitem{BokerHO15}
Udi Boker, Thomas~A. Henzinger, and Jan Otop.
\newblock The target discounted-sum problem.
\newblock In {\em {LICS} Proceedings}, pages 750--761. {IEEE} Computer Society,
  2015.

\bibitem{BouyerBMU15}
Patricia Bouyer, Romain Brenguier, Nicolas Markey, and Michael Ummels.
\newblock Pure {N}ash equilibria in concurrent deterministic games.
\newblock {\em Logical Methods in Comput. Sci.}, 11(2), 2015.

\bibitem{Brenguier16}
Romain Brenguier.
\newblock Robust equilibria in mean-payoff games.
\newblock In {\em Proceedings of {FOSSACS}}, volume 9634 of {\em Lecture Notes
  in Comput. Sci.}, pages 217--233. Springer, 2016.

\bibitem{BrenguierCHPRRS16}
Romain Brenguier, Lorenzo Clemente, Paul Hunter, Guillermo~A. P{\'{e}}rez,
  Mickael Randour, Jean{-}Fran{\c{c}}ois Raskin, Ocan Sankur, and Mathieu
  Sassolas.
\newblock Non-zero sum games for reactive synthesis.
\newblock In {\em {LATA} Proceedings}, volume 9618 of {\em Lecture Notes in
  Comput. Sci.}, pages 3--23. Springer, 2016.

\bibitem{BrenguierRS17}
Romain Brenguier, Jean{-}Fran{\c{c}}ois Raskin, and Ocan Sankur.
\newblock Assume-admissible synthesis.
\newblock {\em Acta Inf.}, 54(1):41--83, 2017.

\bibitem{BBMR15}
Thomas Brihaye, V{\'{e}}ronique Bruy{\`{e}}re, No{\'{e}}mie Meunier, and
  Jean{-}Fran{\c{c}}ois Raskin.
\newblock Weak subgame perfect equilibria and their application to quantitative
  reachability.
\newblock In {\em {CSL} Proceedings}, volume~41 of {\em LIPIcs}, pages
  504--518. Schloss Dagstuhl - Leibniz-Zentrum fuer Informatik, 2015.

\bibitem{TJS13}
Thomas Brihaye, Julie {De Pril}, and Sven Schewe.
\newblock Multiplayer cost games with simple {N}ash equilibria.
\newblock In {\em {LFCS} Proceedings}, volume 7734 of {\em Lecture Notes in
  Comput. Sci.}, pages 59--73. Springer, 2013.

\bibitem{BruyereHR16}
V{\'{e}}ronique Bruy{\`{e}}re, Quentin Hautem, and Jean{-}Fran{\c{c}}ois
  Raskin.
\newblock On the complexity of heterogeneous multidimensional games.
\newblock In {\em {CONCUR} Proceedings}, volume~59 of {\em LIPIcs}, pages
  11:1--11:15. Schloss Dagstuhl - Leibniz-Zentrum fuer Informatik, 2016.

\bibitem{BMR14}
V{\'{e}}ronique Bruy{\`{e}}re, No{\'{e}}mie Meunier, and Jean{-}Fran{\c{c}}ois
  Raskin.
\newblock Secure equilibria in weighted games.
\newblock In {\em {CSL-LICS} Proceedings}, pages 26:1--26:26. {ACM}, 2014.

\bibitem{Bruyere0PR17}
V{\'{e}}ronique Bruy{\`{e}}re, St{\'{e}}phane~Le Roux, Arno Pauly, and
  Jean{-}Fran{\c{c}}ois Raskin.
\newblock On the existence of weak subgame perfect equilibria.
\newblock In {\em {FoSSaCS} Proceedings}, volume 10203 of {\em Lecture Notes in
  Comput. Sci.}, pages 145--161. Springer, 2017.

\bibitem{BuhrkeLV96}
Nils Buhrke, Helmut Lescow, and Jens V{\"{o}}ge.
\newblock Strategy construction in infinite ganes with {S}treett and {R}abin
  chain winning conditions.
\newblock In {\em {TACAS} Proceedings}, volume 1055 of {\em Lecture Notes in
  Comput. Sci.}, pages 207--224. Springer, 1996.

\bibitem{Calude17}
Cristian Calude, Sanjay Jain, Bakhadyr Khoussainov, Wei Li, and Frank Stephan.
\newblock Deciding parity games in quasipolynomial time.
\newblock In {\em {STOC} Proceedings (to appear)}. {ACM}, 2017.

\bibitem{ChatterjeeDFR17}
Krishnendu Chatterjee, Laurent Doyen, Emmanuel Filiot, and
  Jean{-}Fran{\c{c}}ois Raskin.
\newblock Doomsday equilibria for omega-regular games.
\newblock {\em Inf. Comput.}, 254:296--315, 2017.

\bibitem{LaurentDoyen}
Krishnendu Chatterjee, Laurent Doyen, and Thomas~A. Henzinger.
\newblock Quantitative languages.
\newblock {\em ACM Trans. Comput. Log.}, 11, 2010.

\bibitem{Chatterjee0RR15}
Krishnendu Chatterjee, Laurent Doyen, Mickael Randour, and
  Jean{-}Fran{\c{c}}ois Raskin.
\newblock Looking at mean-payoff and total-payoff through windows.
\newblock {\em Inf. Comput.}, 242:25--52, 2015.

\bibitem{ChatterjeeDHL16}
Krishnendu Chatterjee, Wolfgang Dvor{\'{a}}k, Monika Henzinger, and Veronika
  Loitzenbauer.
\newblock Conditionally optimal algorithms for generalized {B}{\"{u}}chi games.
\newblock In {\em {MFCS} Proceedings}, volume~58 of {\em LIPIcs}, pages
  25:1--25:15. Schloss Dagstuhl - Leibniz-Zentrum fuer Informatik, 2016.

\bibitem{ChatterjeeH14}
Krishnendu Chatterjee and Monika Henzinger.
\newblock Efficient and dynamic algorithms for alternating {B}{\"{u}}chi games
  and maximal end-component decomposition.
\newblock {\em J. {ACM}}, 61(3):15:1--15:40, 2014.

\bibitem{CHJ06}
Krishnendu Chatterjee, Thomas~A. Henzinger, and Marcin Jurdzinski.
\newblock Games with secure equilibria.
\newblock {\em Theor. Comput. Sci.}, 365:67--82, 2006.

\bibitem{ChatterjeeHP07}
Krishnendu Chatterjee, Thomas~A. Henzinger, and Nir Piterman.
\newblock Generalized parity games.
\newblock In {\em {FoSSaCS} Proceedings}, volume 4423 of {\em Lecture Notes in
  Comput. Sci.}, pages 153--167. Springer, 2007.

\bibitem{ChatterjeeMJ04}
Krishnendu Chatterjee, Rupak Majumdar, and Marcin Jurdzinski.
\newblock On {N}ash equilibria in stochastic games.
\newblock In {\em {CSL} Proceedings}, volume 3210 of {\em Lecture Notes in
  Comput. Sci.}, pages 26--40. Springer, 2004.

\bibitem{ConduracheFGR16}
Rodica Condurache, Emmanuel Filiot, Raffaella Gentilini, and
  Jean{-}Fran{\c{c}}ois Raskin.
\newblock The complexity of rational synthesis.
\newblock In {\em Proceedings of {ICALP}}, volume~55 of {\em LIPIcs}, pages
  121:1--121:15. Schloss Dagstuhl - Leibniz-Zentrum fuer Informatik, 2016.

\bibitem{DePril13}
Julie {De Pril}.
\newblock {\em Equilibria in Multiplayer Cost Games}.
\newblock PhD thesis, University UMONS, 2013.

\bibitem{Depril14}
Julie {De Pril}, J{\'{a}}nos Flesch, Jeroen Kuipers, Gijs Schoenmakers, and
  Koos Vrieze.
\newblock Existence of secure equilibrium in multi-player games with perfect
  information.
\newblock In {\em {MFCS} Proceedings}, volume 8635 of {\em Lecture Notes in
  Comput. Sci.}, pages 213--225. Springer, 2014.

\bibitem{Emerson85}
E.~Allen Emerson.
\newblock Automata, tableaux and temporal logics (extended abstract).
\newblock In {\em Logics of Programs, Conference, Proceedings}, volume 193 of
  {\em Lecture Notes in Comput. Sci.}, pages 79--88. Springer, 1985.

\bibitem{EJ91}
E.~Allen Emerson and Charanjit~S. Jutla.
\newblock Tree automata, mu-calculus and determinacy.
\newblock In {\em {FOCS} Proceedings}, pages 368--377. IEEE Comp. Soc., 1991.

\bibitem{EmersonL87}
E.~Allen Emerson and Chin{-}Laung Lei.
\newblock Modalities for model checking: Branching time logic strikes back.
\newblock {\em Sci. Comput. Program.}, 8(3):275--306, 1987.

\bibitem{FijalkowH13}
Nathana{\"{e}}l Fijalkow and Florian Horn.
\newblock Les jeux d'accessibilit{\'{e}} g{\'{e}}n{\'{e}}ralis{\'{e}}e.
\newblock {\em Technique et Science Informatiques}, 32(9-10):931--949, 2013.

\bibitem{filar1997}
Jerzy Filar and Koos Vrieze.
\newblock {\em Competitive {M}arkov Decision Processes}.
\newblock Springer, 1997.

\bibitem{Flesch10}
J{\'{a}}nos Flesch, Jeroen Kuipers, Ayala Mashiah{-}Yaakovi, Gijs Schoenmakers,
  Eilon Solan, and Koos Vrieze.
\newblock Perfect-information games with lower-semicontinuous payoffs.
\newblock {\em Math. Oper. Res.}, 35:742--755, 2010.

\bibitem{Fudenberg83}
Drew Fudenberg and David Levine.
\newblock Subgame-perfect equilibria of finite- and infinite-horizon games.
\newblock {\em Journal of Economic Theory}, 31:251--268, 1983.

\bibitem{GimbertZ04}
Hugo Gimbert and Wieslaw Zielonka.
\newblock When can you play positionally?
\newblock In {\em {MFCS} Proceedings}, volume 3153 of {\em Lecture Notes in
  Comput. Sci.}, pages 686--697. Springer, 2004.

\bibitem{GimbertZ05}
Hugo Gimbert and Wieslaw Zielonka.
\newblock Games where you can play optimally without any memory.
\newblock In {\em {CONCUR} Proceedings}, volume 3653 of {\em Lecture Notes in
  Comput. Sci.}, pages 428--442. Springer, 2005.

\bibitem{2001automata}
Erich Gr{\"{a}}del, Wolfgang Thomas, and Thomas Wilke, editors.
\newblock {\em Automata, Logics, and Infinite Games: {A} Guide to Current
  Research}, volume 2500 of {\em Lecture Notes in Comput. Sci.} Springer, 2002.

\bibitem{GU08}
Erich Gr{\"a}del and Michael Ummels.
\newblock {Solution Concepts and Algorithms for Infinite Multiplayer Games}.
\newblock In {\em {New Perspectives on Games and Interaction}}, volume~4, pages
  151--178. Amsterdam University Press, 2008.

\bibitem{Harris85}
Christopher Harris.
\newblock Existence and characterization of perfect equilibrium in games of
  perfect information.
\newblock {\em Econometrica}, 53:613--628, 1985.

\bibitem{Horn08}
Florian Horn.
\newblock Explicit {M}uller games are {PTIME}.
\newblock In {\em {FSTTCS} Proceedings}, volume~2 of {\em LIPIcs}, pages
  235--243. Schloss Dagstuhl - Leibniz-Zentrum fuer Informatik, 2008.

\bibitem{HunterD05}
Paul Hunter and Anuj Dawar.
\newblock Complexity bounds for regular games.
\newblock In {\em {MFCS} Proceedings}, volume 3618 of {\em Lecture Notes in
  Comput. Sci.}, pages 495--506. Springer, 2005.

\bibitem{HunterR14}
Paul Hunter and Jean{-}Fran{\c{c}}ois Raskin.
\newblock Quantitative games with interval objectives.
\newblock In {\em {FSTTCS} Proceedings}, volume~29 of {\em LIPIcs}, pages
  365--377. Schloss Dagstuhl - Leibniz-Zentrum fuer Informatik, 2014.

\bibitem{Immerman81}
Neil Immerman.
\newblock Number of quantifiers is better than number of tape cells.
\newblock {\em J. Comput. Syst. Sci.}, 22:384--406, 1981.

\bibitem{Jones75}
Neil~D. Jones.
\newblock Space-bounded reducibility among combinatorial problems.
\newblock {\em J. Computer and System Science}, 11:68--75, 1975.

\bibitem{Jurdzinski98}
Marcin Jurdzinski.
\newblock Deciding the winner in parity games is in {UP} $\cap$ co-{UP}.
\newblock {\em Inf. Process. Lett.}, 68(3):119--124, 1998.

\bibitem{karp}
Richard~M Karp.
\newblock A characterization of the minimum cycle mean in a digraph.
\newblock {\em Discrete Mathematics}, 23:309--311, 1978.

\bibitem{Kopczynski06}
Eryk Kopczynski.
\newblock Half-positional determinacy of infinite games.
\newblock In {\em {ICALP} Proceedings}, volume 4052 of {\em Lecture Notes in
  Comput. Sci.}, pages 336--347. Springer, 2006.

\bibitem{kuhn53}
Harold~W. Kuhn.
\newblock {Extensive games and the problem of information}.
\newblock {\em Classics in Game Theory}, pages 46--68, 1953.

\bibitem{Lothaire}
M.~Lothaire.
\newblock {\em Algebraic Combinatorics on Words}, volume~90.
\newblock Cambridge University Press, 2002.

\bibitem{Mar75}
Donald~A. Martin.
\newblock Borel determinacy.
\newblock {\em Annals of Mathematics}, 102:363--371, 1975.

\bibitem{McNaughton93}
Robert McNaughton.
\newblock Infinite games played on finite graphs.
\newblock {\em Ann. Pure Appl. Logic}, 65(2):149--184, 1993.

\bibitem{Nas50}
John~F. Nash.
\newblock Equilibrium points in $n$-person games.
\newblock In {\em PNAS}, volume~36, pages 48--49. National Academy of Sciences,
  1950.

\bibitem{Osborne94}
Martin~J. Osborne and Ariel Rubinstein.
\newblock {\em A course in Game Theory}.
\newblock MIT Press, Cambridge, MA, 1994.

\bibitem{PP04}
Dominique Perrin and Jean-Eric Pin.
\newblock {\em Infinite Words, Automata, Semigroups, Logic and Games}, volume
  141.
\newblock Elsevier, 2004.

\bibitem{PitermanP06}
Nir Piterman and Amir Pnueli.
\newblock Faster solutions of {R}abin and {S}treett games.
\newblock In {\em {LICS} Proceedings}, pages 275--284. {IEEE} Computer Society,
  2006.

\bibitem{Purves11}
Roger~A. Purves and William~D. Sudderth.
\newblock Perfect information games with upper semicontinuous payoffs.
\newblock {\em Math. Oper. Res.}, 36(3):468--473, 2011.

\bibitem{Renyi57}
A.~R\'enyi.
\newblock Representations of real numbers and their ergodic properties.
\newblock {\em Acta Mathematica Academiae Scientiarum Hungarica},
  8(3-4):477--493, 1957.

\bibitem{RouxP16}
St{\'{e}}phane~Le Roux and Arno Pauly.
\newblock Extending finite memory determinacy to multiplayer games.
\newblock In {\em Proceedings of {SR}}, volume 218 of {\em {EPTCS}}, pages
  27--40, 2016.

\bibitem{SafraV89}
Shmuel Safra and Moshe~Y. Vardi.
\newblock On omega-automata and temporal logic (preliminary report).
\newblock In {\em Proceedings of {STOC}}, pages 127--137. {ACM}, 1989.

\bibitem{selten}
Reinhard Selten.
\newblock Spieltheoretische {B}ehandlung eines {O}ligopolmodells mit
  {N}achfragetr\"agheit.
\newblock {\em Zeitschrift f\"ur die gesamte Staatswissenschaft}, 121:301--324
  and 667--689, 1965.

\bibitem{SV03}
Eilon Solan and Nicolas Vieille.
\newblock Deterministic multi-player {D}ynkin games.
\newblock {\em Journal of Mathematical Economics}, 39:911--929, 2003.

\bibitem{Ummels06}
Michael Ummels.
\newblock Rational behaviour and strategy construction in infinite multiplayer
  games.
\newblock In {\em {FSTTCS} Proceedings}, volume 4337 of {\em Lecture Notes in
  Comput. Sci.}, pages 212--223. Springer, 2006.

\bibitem{Ummels08}
Michael Ummels.
\newblock The complexity of {N}ash equilibria in infinite multiplayer games.
\newblock In {\em {FoSSaCS} Proceedings}, volume 4962 of {\em Lecture Notes in
  Comput. Sci.}, pages 20--34. Springer, 2008.

\bibitem{UmmelsW11}
Michael Ummels and Dominik Wojtczak.
\newblock The complexity of {N}ash equilibria in limit-average games.
\newblock In {\em {CONCUR} Proceedings}, volume 6901 of {\em Lecture Notes in
  Comput. Sci.}, pages 482--496. Springer, 2011.

\bibitem{UmmelsArXiv}
Michael Ummels and Dominik Wojtczak.
\newblock The complexity of {N}ash equilibria in limit-average games.
\newblock {\em CoRR}, abs/1109.6220, 2011.

\bibitem{VardiW94}
Moshe~Y. Vardi and Pierre Wolper.
\newblock Reasoning about infinite computations.
\newblock {\em Inf. Comput.}, 115(1):1--37, 1994.

\bibitem{Velner15}
Yaron Velner.
\newblock Robust multidimensional mean-payoff games are undecidable.
\newblock In {\em FoSSaCS Proceedings}, volume 9034 of {\em Lecture Notes in
  Comput. Sci.}, pages 312--327. Springer, 2015.

\bibitem{VelnerC0HRR15}
Yaron Velner, Krishnendu Chatterjee, Laurent Doyen, Thomas~A. Henzinger,
  Alexander~Moshe Rabinovich, and Jean{-}Fran{\c{c}}ois Raskin.
\newblock The complexity of multi-mean-payoff and multi-energy games.
\newblock {\em Inf. Comput.}, 241:177--196, 2015.

\bibitem{vonNeumannMorgenstern44}
John von Neumann and Oskar Morgenstern.
\newblock {\em Theory of Games and Economic Behavior}.
\newblock Princeton University Press, 1944.

\bibitem{ZP96}
Uri Zwick and Mike Paterson.
\newblock The complexity of mean payoff games on graphs.
\newblock {\em Theor. Comput. Sci.}, 158:343--359, 1996.

\end{thebibliography}

\section*{Appendix}
In this appendix, we give a sketch of proof for Muller games in Theorem~\ref{thm:NEprob2}. Recall that each player~$i$ has the objective $\Omega_i = \{\rho \in \Plays \mid \Infrho(\rho) \in {\cal F}_i \}$ with ${\cal F}_i \subseteq 2^V$, and that the values $val_i(v)$, $v \in V$, in each game $G_i$ can be computed in polynomial time (Theorem~\ref{thm:TwoPlayerOmegaReg}). To prove {\sf P} membership for the constraint problem with bounds $(\mu_i)_{i \in \Pi}, (\nu_i)_{i \in \Pi}$, we apply the approach (\ref{eq:constraint}). Notice that for the required play $\rho \in \Plays(v_0)$ in (\ref{eq:constraint}), the set $U = \Infrho(\rho)$ must be a strongly connected component that is reachable from the initial vertex $v_0$. 
Moreover if for some $i$, $f_i(\rho) = 0$ then $\val_i(\rho_k) = 0$ for all $\rho_k \in V_i$, and if $\nu_i = 0$, then $f_i(\rho) = 0$. 
We thus proceed as follows. \emph{(i)} For each $i$ such that $\nu_i = 1$, for each $U \in {\cal F}_i$ (seen as a potential $U = \Infrho(\rho)$), the following computations are done in polynomial time for all $j\in\Pi$:
\begin{itemize}
\item if $\mu_j = 1$ ((\ref{eq:constraint}) imposes $f_j(\rho) = 1$), test whether $U \in {\cal F}_j$,
\item if $\nu_j = 0$ ((\ref{eq:constraint}) imposes $f_j(\rho) = 0$), test whether $U \not\in {\cal F}_j$ and whether each $v \in U \cap V_j$ has value $\val_j(v) = 0$,
\item if $\mu_j = 0$ and $\nu_j = 1$ ((\ref{eq:constraint}) allows either $f_j(\rho) = 0$ or $f_j(\rho) = 1$), then if $U \not\in {\cal F}_j$, test whether each $v \in U \cap V_j$ has value $\val_j(v) = 0$.
\end{itemize}
Finally, construct in polynomial time the game $G'$ from $G$ such that each $V_j$ is limited to $\{v \in V_j \mid \val_j(v) = 0 \}$ whenever $U \not\in {\cal F}_j$, and test whether $U$ is a strongly connected component that is reachable from $v_0$ in $G'$.
As soon as this sequence of tests is positive, there exists $\rho$ satisfying (\ref{eq:constraint}). 
\emph{(ii)} It may happen that step \emph{(i)} cannot be applied (because there is no $j$ such that $\mu_j = 1$, and for $j$ such that $\mu_j = 0$ and $\nu_j = 1$, there is no potential $U = \Infrho(\rho)$). In this case, we construct in polynomial time a two-player game $G'$ from $G$ such that each $V_i$ is limited to $\{v \in V_i \mid \val_i(v) = 0 \}$, player~$1$ controls no vertex and player~$2$ is formed by the coalition of all $i \in \Pi$, and the objective is a Muller objective with ${\cal F} = \cup_{i\in \Pi} {\cal F}_i$. We then test in polynomial time whether player~$1$ has no winning strategy from $v_0$ in this Muller game. 

\end{document}